\documentclass[11pt,prd,preprintnumbers,superscriptaddress,nofootinbib]{revtex4}
\usepackage{graphicx}
\usepackage{dcolumn}
\usepackage{bm}
\usepackage{amssymb}
\usepackage{amsmath}
\usepackage{epsfig}
\usepackage{hhline}
\usepackage{color}
\usepackage{multirow}
\usepackage{cancel}
\usepackage[normalem]{ulem}
\usepackage[colorlinks,citecolor=blue]{hyperref}
\usepackage{enumitem}

\def\lapp{\mathrel{\rlap{\raise.5ex\hbox{$<$}}
                    {\lower.5ex\hbox{$\sim$}}}}
\def\gapp{\mathrel{\rlap{\raise.5ex\hbox{$>$}}
                    {\lower.5ex\hbox{$\sim$}}}}

{
{

\newcommand{\bmt}{\begin{pmatrix}}
\newcommand{\emt}{\end{pmatrix}}
\newcommand{\ba}{\begin{array}{c}}
\newcommand{\ea}{\end{array}}
\newcommand{\be}{\begin{equation}}
\newcommand{\ee}{\end{equation}}
\newcommand{\bea}{\begin{eqnarray}}
\newcommand{\eea}{\end{eqnarray}}

\newcommand{\bi}{\begin{itemize}}
\newcommand{\ei}{\end{itemize}}

\newcommand{\baz}{\begin{array}{cc}}

\newcommand{\mathsym}[1]{{}}

\newcommand{\bt}{\begin{tabular}}
\newcommand{\et}{\end{tabular}}

\newcommand{\benu}{\begin{enumerate}}
\newcommand{\eenu}{\end{enumerate}}
\newcommand{\bav}{\begin{array}{cccc}}
\begin{document}
\title{\bf Flavoured CP-asymmetry at the effective neutrino mass floor}
\author{Nimmala Narendra}
\email{ph14resch01002@iith.ac.in}
\author{Narendra Sahu}
\email{nsahu@iith.ac.in}
\affiliation{Department of Physics, Indian Institute of Technology Hyderabad, Kandi, Sangareddy, 502285, Telangana, India.}
\author{S. Uma Sankar}
\email{uma@phy.iitb.ac.in}
\affiliation{ Department of Physics, Indian Institute of Technology Bombay,
 \\  Powai, Mumbai 400076, India.}

\begin{abstract}
Both neutrinoless double beta decay and leptogenesis require neutrinos to be Majorana fermions. A relation between these two phenomena can be derived once the mechanism of neutrino mass generation is specified. We first derive the constraints on the  Majorana phases by minimising the effective neutrino mass in neutrinoless double beta decay with respect to the smallest mass among the light neutrinos.  Given these phases, we derive a lower bound on $M_{1}$ (the mass of the lightest of the heavy neutrinos) in the framework of Type-I seesaw mechanism, subject to the constraint that the CP asymmetry required for adequate leptogenesis is larger than $10^{-8}$. We find that $M_{1} \geq 10^{10}\,(10^{9})$ GeV for the case of Normal (Inverted) hierarchy. We extend our analysis to the case when one of the heavy neutrinos decouples (two right handed neutrino models). In this case we find $M_{1} \geq 10^{10}\,(10^{11})$ GeV for the case of Normal (Inverted) hierarchy.  
\end{abstract}
\maketitle
\newpage
\section{Introduction} \label{Intro}
At present the Standard Model (SM) of particle physics, which is based on the  gauge group $SU(3)_C \times SU(2)_L \times 
U(1)_Y$, is considered to be the best candidate to explain elementary particles and their interactions in nature. However, it 
doesn't address certain issues like sub-eV masses of three generations of active neutrinos. Moreover, it does not explain the 
observed baryon asymmetry of the Universe, measured to be $n_B/n_\gamma = (6.09 \pm 0.06)\times 10^{-10}$ \cite{Hinshaw:2012aka,Aghanim:2018eyx}. Therefore, we need to go beyond the SM of particle physics to address these issues.  

The current neutrino oscillation experiments~\cite{solar-expt, atmos-expt} confirmed non-zero, but tiny neutrino masses and also mixing between different flavours. The mixing matrix, relating the flavour eigenstates to mass eigenstates, is parameterized as~\cite{pmns-matrix}   
\begin{equation}
U_{\text{PMNS}}=\left(\begin{array}{ccc}
c_{12}c_{13}& s_{12}c_{13}& s_{13}e^{-i\delta}\\
-s_{12}c_{23}-c_{12}s_{23}s_{13}e^{i\delta}& c_{12}c_{23}-s_{12}s_{23}s_{13}e^{i\delta} & s_{23}c_{13} \\
s_{12}s_{23}-c_{12}c_{23}s_{13}e^{i\delta} & -c_{12}s_{23}-s_{12}c_{23}s_{13}e^{i\delta}& c_{23}c_{13}
\end{array}\right) U_{ph}\, , 
\label{matrixUPMNS}
\end{equation}
where $U_{ph}={\rm diag}(1, e^{i \alpha_{1}}, e^{i \alpha_{2}})$ with $\alpha_{1}$, $\alpha_{2}$ being the Majorana phases and 
$\delta$ is Dirac phase. The symbols $c_{ij}$ and $s_{ij}$ stand for $\cos{\theta_{ij}}$ and $\sin{\theta_{ij}}$ respectively.

In terms of the mixing matrix $U_{\text{PMNS}}$ the neutrino mass matrix can be given as 
\begin{equation}
m_{\nu}= U^{\dagger} m_{\rm diag} U^{*}
\end{equation}
where $m_{\rm diag}={\rm diag}(m_1, m_2, m_3)$. Thus the neutrino mass matrix $m_\nu$ consists of nine parameters: three masses, three mixing angles and three phases. At present the oscillation experiments measure two mass square differences: namely solar ($\Delta m^{2}_{\rm sol}$) and atmospheric ($\Delta m^{2}_{\rm atm}$), three mixing angles $\theta_{23},\, \theta_{12}$ and $\theta_{13}$ to a good degree of precision. Data indicate that $|\Delta m_{\rm atm}^{2}| \gg \Delta m_{{\rm sol}}^{2}$. Without loss of generality, we can define $\Delta m_{{\rm sol}}^{2}= \Delta m_{{\rm 21}}^{2}= m_{2}^{2}-m_{1}^{2}$ and $\Delta m_{{\rm atm}}^{2}=\Delta m_{{\rm 31}}^{2}= m_{3}^{2}-m_{1}^{2}\simeq m_{3}^{2}-m_{2}^{2}=\Delta m_{{\rm 32}}^{2}$. Matter effects in solar neutrino oscillations require $\Delta m_{{\rm 21}}^{2} > 0$, but, so far, the sign of  $\Delta m_{{\rm 31}}^{2}$ is not determined. The case of $\Delta m_{31}^{2}>0$ is called Normal heirarchy (NH) and that of $\Delta m_{31}^{2}<0$ is called Inverted hierarchy (IH). For NH the smallest neutrino mass ($m_{\rm min}$) is $m_{1}$, where as it is $m_{3}$ for IH. The 3$\sigma$ ranges of the oscillation parameters are given in Table~\ref{tabglobalfit}. 
\begin{table}[htb]
\centering
\begin{tabular}{|c|c|c|}
\hline
Parameters & Normal Hierarchy (NH) & Inverted Hierarchy (IH) \\
\hline
$ \frac{\Delta m_{21}^2}{10^{-5} \text{eV}^2}$ & $6.79-8.01$ & $6.79-8.01 $ \\
$ \frac{|\Delta m_{31}^2|}{10^{-3} \text{eV}^2}$ & $2.427-2.625$ & $2.412-2.611 $ \\
$ \sin^2\theta_{12} $ &  $0.275-0.350 $ & $0.275-0.350 $ \\
$ \sin^2\theta_{23} $ & $0.418-0.627$ &  $0.423-0.629 $ \\
$\sin^2\theta_{13} $ & $0.02045-0.02439$ & $0.02068-0.02463 $ \\
$ \delta (^\circ) $ & $125-392$ & $196-360$ \\
\hline
\end{tabular}
\caption{Global fit $3\sigma$ ranges of neutrino oscillation parameters \cite{Esteban:2018azc}.}
\label{tabglobalfit}
\end{table}

At present the value of Dirac phase $\delta$ is quite ambiguous. T2K experiment\,\cite{Abe:2011sj} prefers a value $\delta \approx -\pi/2$ whereas NOvA experiment\,\cite{Acero:2019ksn} prefers $\delta \approx 0$. The best fit values for global fits is $\delta \approx - 3 \pi/4$ for NH and $\delta \approx - \pi/2$ for IH\,\cite{Esteban:2018azc}. 

The oscillation experiments do not give us any hint towards the nature of neutrino to be either Dirac or Majorana. However, 
the neutrinoless double beta decay ($0\nu\beta \beta$) experiments~\cite{Agostini:2017iyd} can explore the Majorana nature of neutrinos. Till date the best lower limit on half-life of the $0\nu\beta \beta$ using $^{76}$Ge is $T_{1/2}^{0\nu} > 8.0 \times 10^{25}$ yrs at 90\% C.L. from GERDA~\cite{Agostini:2018tnm}. For $^{136}$Xe isotope, the derived lower limits on half-life from KamLAND-Zen experiment is $T_{1/2}^{0\nu} > 1.6 \times 10^{26}$ yrs~\cite{KamLAND-Zen:2016pfg}. The proposed sensitivity of the future planned nEXO experiment is  $T_{1/2}^{0\nu} \approx 6.6 \times 10^{27}$ yrs~\cite{Albert:2014awa}. The above mentioned lower limits on $T_{1/2}^{0\nu}$ lead to an upper limit on effective neutrino mass $|m_{ee}| \approx 10^{-2}$ eV. In future these experiments will increase their sensitivities down to the floor of $|m_{ee}|$, i.e. the minimum of $|m_{ee}|$ for all allowed values of $m_{\rm min}$.

Since Majorana neutrinos violate lepton number by two units they can lead to leptogenesis. Within a given neutrino mass model, we can expect a relation between neutrinoless double beta decay and leptogenesis. The elegant type-I seesaw mechanism~\cite{Minkowski:1977sc, GellMann:1980vs, type1_seesaw, Mohapatra:1979ia, Schechter:1980gr, Ma:1998dn, Magg:1980ut,Cheng:1980qt,Gelmini:1980re, Ma:1998dx} requires only the addition of three right handed neutrinos $N_i$ $(i=1,2,3)$ to the SM. Since these particles are electrically neutral and have no charges under the SM gauge group, they can have Majorana masses $M_i$ $(i=1,2,3)$. The CP-violating out-of-equilibrium decay of the lightest of these heavy neutrinos ($N_{1}$) in the early Universe could generate a net lepton asymmetry, which is then converted to observed baryon asymmetry of the Universe by electroweak sphalerons\,\cite{Fukugita:1986hr, Arnold:1987mh}. 

In this work, we first minimise $|m_{ee}|$ as function of $m_{\rm min}$ and to find the ranges of values of Majorana phases $\alpha_{1}$ and $\alpha_{2}$ which yield the minimum value. With these phases as inputs, we compute the CP-asymmetries $\epsilon_{1}^{l}$ ($l=e, \mu, \tau$) responsible for leptogenesis\,\cite{sakharov.67,Fukugita:1986hr,baryo_lepto_group, Abada:2006ea, Pilaftsis:2003gt, Pilaftsis:2005rv, 2RH-models, rebelo_prd}. We derive a lower bound on $M_{1}$ (mass of $N_{1}$) both in case of three\,\cite{3RH-models, Branco:2006ce, scpv_models} and two right handed neutrino scenarios\,\cite{Asaka:2018hyk, Bhattacharya:2006aw} by imposing the constraint $\epsilon_{1}^{l} \geq 10^{-8}$, which can give rise adequate leptogenesis.

The paper is organised as follows. In the section\,\ref{eff_maj_mass}, we minimise the effective neutrino mass parameter $|m_{ee}|$ 
as a function of $m_{\rm min}$ and obtain the allowed range of Majorana phases. We then obtain the flavour dependent CP-asymmetry parameters in section\,\ref{CP-asy_3RHN} corresponding to the set of low energy parameters which minimise $|m_{ee}|$. In section\,\ref{CP-asy_2RHN}. We repeat this calculation for the case where there are only two heavy right handed neutrinos (which corresponds to setting $m_{\rm min}=0$). We present our conclusion in the last section\,\ref{conclusion}.  

\section{Effective Majorana Mass and its minimisation}\label{eff_maj_mass}
The most promising way to find the Majorana nature of neutrinos is through $0\nu \beta \beta$ decay experiments. The $0\nu \beta \beta$ decay rate is proportional to the effective Majorana mass $|m_{ee}|$, which is given by,
\begin{eqnarray}
|m_{ee}|&=&|m_{1} \cos^{2}\theta_{12} \cos^{2}\theta_{13}+m_{2}\cos^{2}\theta_{13} \sin^{2}\theta_{12} e^{-2 i \alpha_{1}}+m_{3}\sin^{2}\theta_{13} e^{-2 i (\alpha_{2}-\delta)}|.
\label{mee_expr}
\end{eqnarray}
As we see from Eq.\,\ref{mee_expr}, the value of effective Majorana mass depends on light neutrino masses $m_1, m_2, m_3$, mixing angles $\theta_{12}, \theta_{13}$ and three CP-phases $\delta$, $\alpha_1$ and $\alpha_2$. Given the minimum mass ($m_{\rm min}$), the other two neutrino masses can be defined in terms of the $\Delta m^{2}_{\rm 21}$ and the $\Delta m^{2}_{\rm 31}$. For NH, $m_{1}$ is $m_{\rm min}$ and $m_2 = \sqrt{m_{\rm min}^{2}+\Delta m_{\rm 21}^{2}},\, m_3 = \sqrt{m_{\rm min}^{2}+\Delta m_{\rm 31}^{2}}$. For IH, $m_{3}$ is $m_{\rm min}$ and $\Delta m^{2}_{\rm 31}$ is negative. The expressions for the other two masses are $m_{1}= \sqrt{m_{\rm min}^{2}+|\Delta m_{\rm 31}^{2}|}$, $m_{2}=\sqrt{m_{\rm min}^{2}+|\Delta m_{\rm 31}^{2}|+ \Delta m^{2}_{\rm 21}}$. At present, the three mixing angles are strongly constrained by the neutrino oscillation data, while the three phases, which are responsible for CP-violation\,\cite{dirac_phase_group}, are essentially unconstrained. 

First, we consider the dependence of the minimum of $|m_{ee}|$ on $m_{\rm min}$ qualitatively based on Eq.\,\ref{mee_expr}. Here we take the ranges of the three phases to be $(-\pi,+\pi)$. We need to consider the cases of NH and IH seperately. For NH, we consider four different mass sub-ranges for $m_{\rm min}=m_{1}$. 
\begin{enumerate}[label=\roman*).]
\item \underline{$m_{1}$, $(10^{-2} - 1)\, {\rm eV}$:} In this case $m_{1}\simeq m_{2} \simeq m_{3}$. The first term in Eq.\,\ref{mee_expr} is positive and has the largest magnitude. To obtain the minimum of $|m_{ee}|$ the second and third term have to be negative. This leads to the conditions $\alpha_{1}=\pm \pi/2$ and $\alpha_{2}=\delta \pm \pi/2$. 
\item \underline{$m_{1}$, $(10^{-3} - 10^{-2})\, {\rm eV}$:} Here $m_{2} \gtrsim 0.01$ eV and $m_{3} \simeq 0.05$ eV. Given $\sin^{2}\theta_{12} \approx 0.35$ and $\sin^{2}\theta_{13} \approx 0.02$, the $|m_{ee}|$ is dominated by the second term and the third term is negligibly small. Hence the minimum of $|m_{ee}|$ occurs for $\alpha_{1} \approx \pm \pi/2$, and depends very weakly on $\alpha_{2}$ and $\delta$. In this case, the magnitudes of $m_{1}$ and $m_{2}$ are comparable and hence the complete cancellations between the first two terms of Eq.\,\ref{mee_expr} is possible. In such a situation, it is possible for $|m_{ee}| \rightarrow 0$.  
\item \underline{$m_{1}$, $(10^{-4} - 10^{-3})\, {\rm eV}$:} Here again the second term of Eq.\,\ref{mee_expr} has the largest magnitude with the first and third having similar magnitudes. Requiring both these to have sign opposite to that of the second term imposes the conditions $\alpha_{1}=\pm \pi/2$ and $\alpha_{2} \approx \delta \pm \pi$. The minimisation procedure leads to a lower bound on $|m_{ee}| \simeq 10^{-3}$ eV.
\item \underline{$m_{1}$, $(10^{-6} - 10^{-4})\, {\rm eV}$:} This case is similar to the third case except that the first term is negligibly small. The minimisation condition is equivalent to the requirement that the terms two and three must have opposite sign. This occurs when $\alpha_{2}=\alpha_{1}+\delta - (2n+1)\pi/2$ for appropriate integer values of $n$. Here also we obtain a lower bound on $|m_{ee}| \simeq 10^{-3}$ eV.
\end{enumerate}
Turning to the case of IH, we consider the following two sub-ranges for $m_{\rm min}=m_{3}$.
\begin{enumerate}[label=\roman*).]
\item \underline{$m_{3}$, $(10^{-6} - 10^{-2})\, {\rm eV}$:} This case similar to first case of NH. Hence the conditions on $\alpha_{1}$ and $\alpha_{2}$ are same. 
\item \underline{$m_{3}$, $(10^{-2} - 1)\, {\rm eV}$:} The smallness of $m_{3}$ and $\sin^{2}\theta_{13}$ make the third term in Eq.\,\ref{mee_expr} completely negligible and the minimization of $|m_{ee}|$ completely independent of $\alpha_{2}$ and $\delta$. Requiring a cancellation between the first two terms, we get $\alpha_{1}=\pm \pi/2$.  
\end{enumerate}

\begin{figure} [htbp]
	\centering
	\includegraphics[width=80mm]{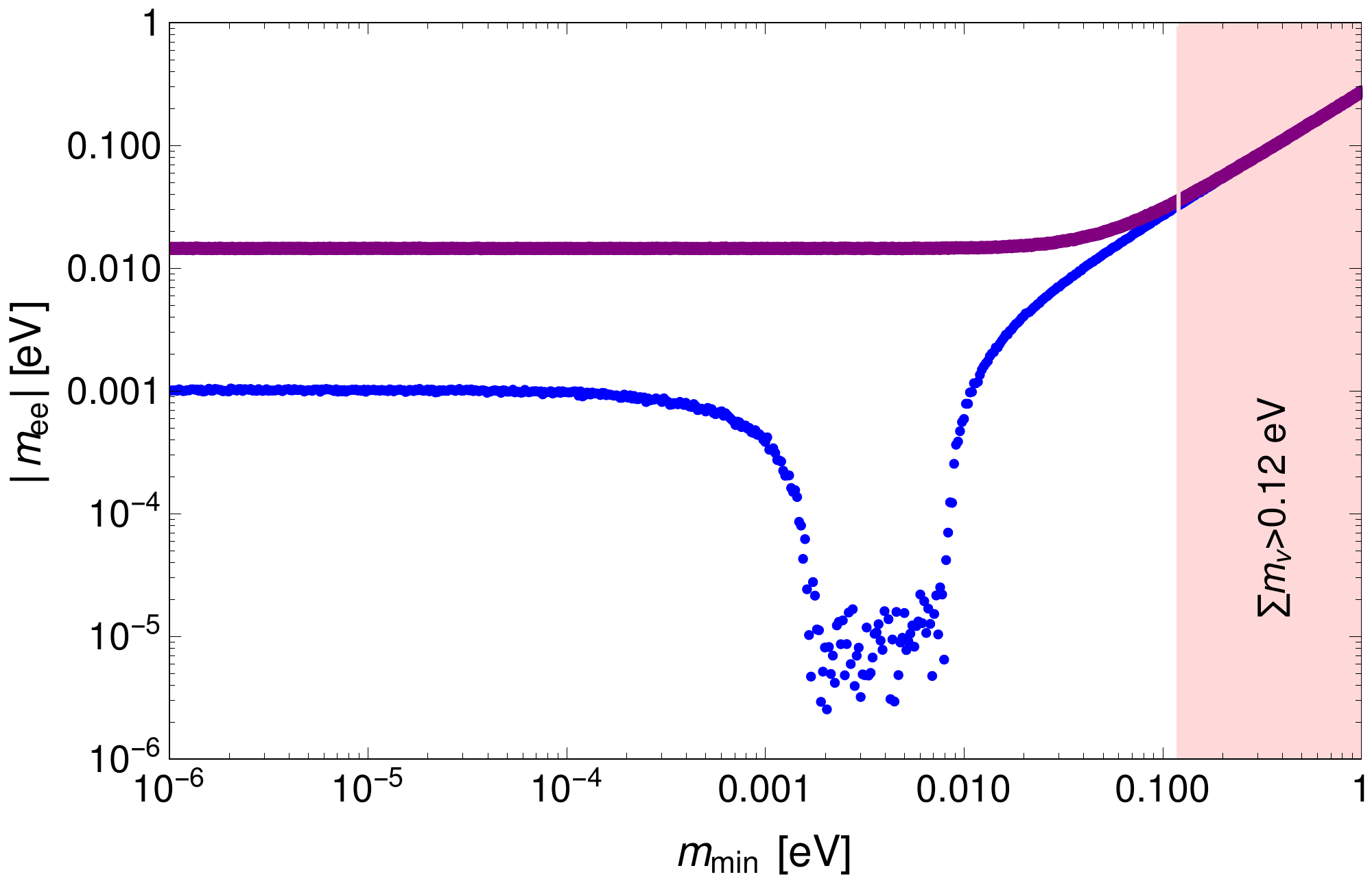}\\
   \caption{\footnotesize{Minimum of $|m_{ee}|$ as a function of $m_{\rm min}$ in case of normal hierarchy (Blue) and inverted hierarchy (Purple). The Pink shaded region defined by $m_{\rm min} > 0.12 {\rm eV}$ is ruled out by PLANCK data~\cite{Aghanim:2018eyx}.}}
	\label{mmin_vs_mee}
\end{figure}

To verify the qualitative deductions made above, we calculated the minimum of $|m_{ee}|$ as a function of $m_{\rm min}$ through simulations. We varied $m_{\rm min}$ within its sub-range, the neutrino oscillation parameters within their 3$\sigma$ ranges and the two Majorana phases $\alpha_1$ and $\alpha_2$ in $(-\pi$, $+\pi)$ while keeping the Dirac phase fixed at $\delta$= -$\pi/2$, which is close to the current best fit of global data\,\cite{Esteban:2018azc}.  We randomly chose a set of values of the neutrino parameters within their respective ranges and calculated $|m_{ee}|$. We repeated this procedure $10^{7}$ times and picked the minimum value of $|m_{ee}|$ and the corresponding values of $\alpha_{1}$ and $\alpha_{2}$. The variation of $|m_{ee}|$ with respect to $m_{\rm min}$ as shown in Fig.\,\ref{mmin_vs_mee}, where  $|m_{ee}| = 10^{-3} \,(10^{-2})$ eV for NH (IH) as $m_{\rm min} \rightarrow 0$. We note that $|m_{ee}|\rightarrow 0$ for the case of NH if $m_{\rm min}$ is in the sub-range $(10^{-3}-10^{-2})$ eV \,\cite{Bilenky:2014uka, majorana_phase_group}. 

\begin{figure}
{{\includegraphics[width=70mm]{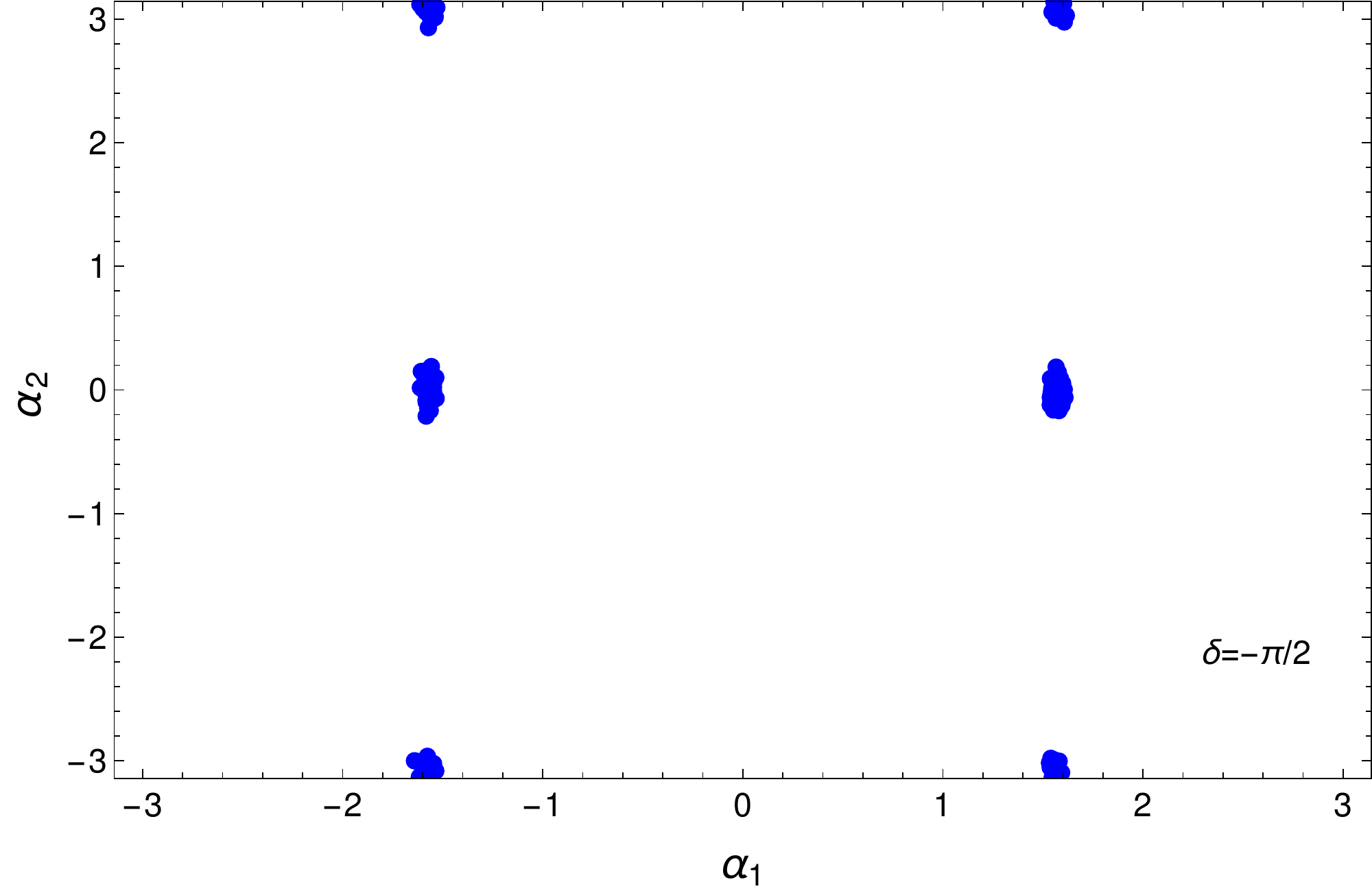}}}
 	     \qquad
{{\includegraphics[width=70mm]{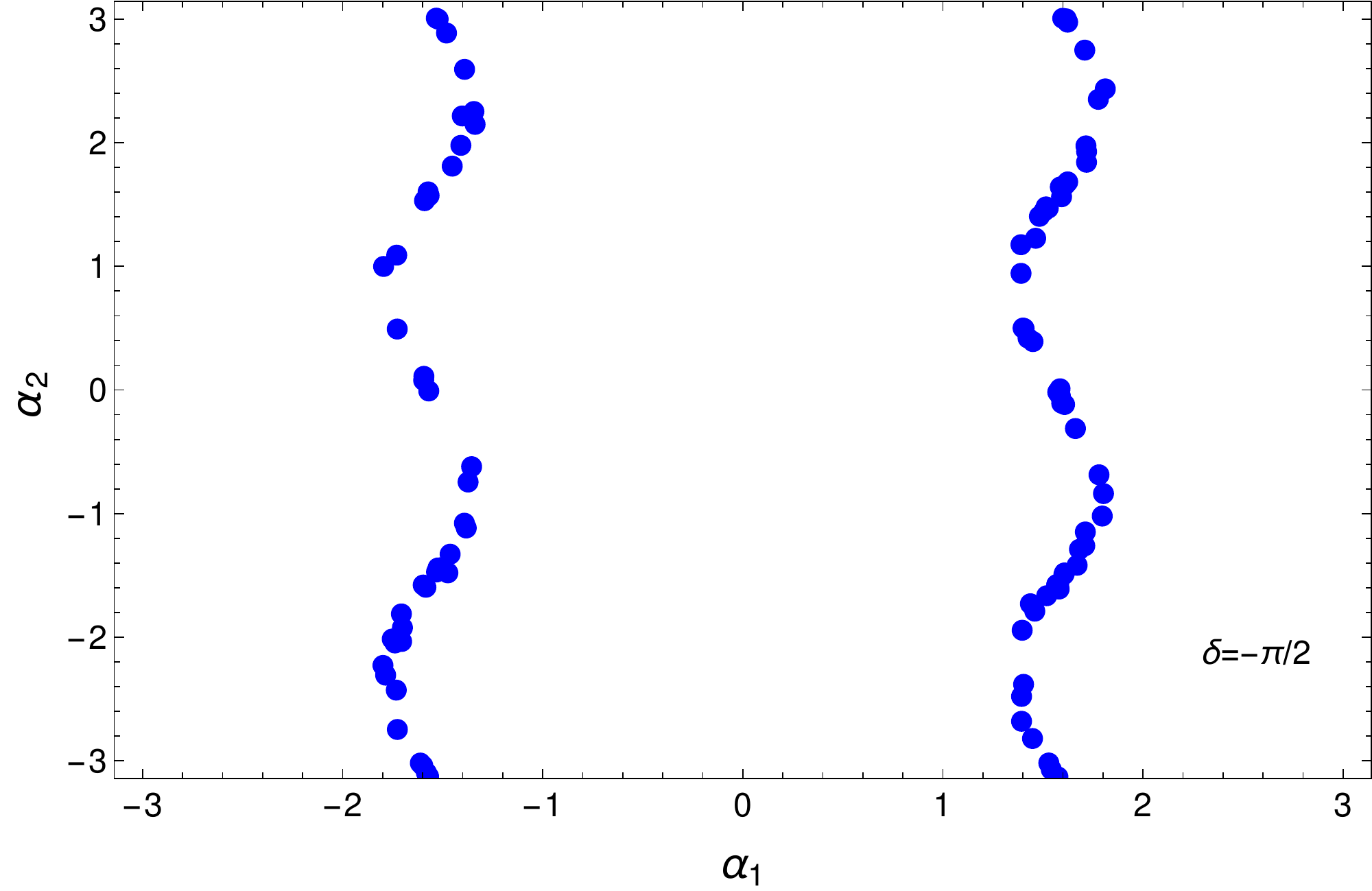}}}
	     \qquad
{{\includegraphics[width=70mm]{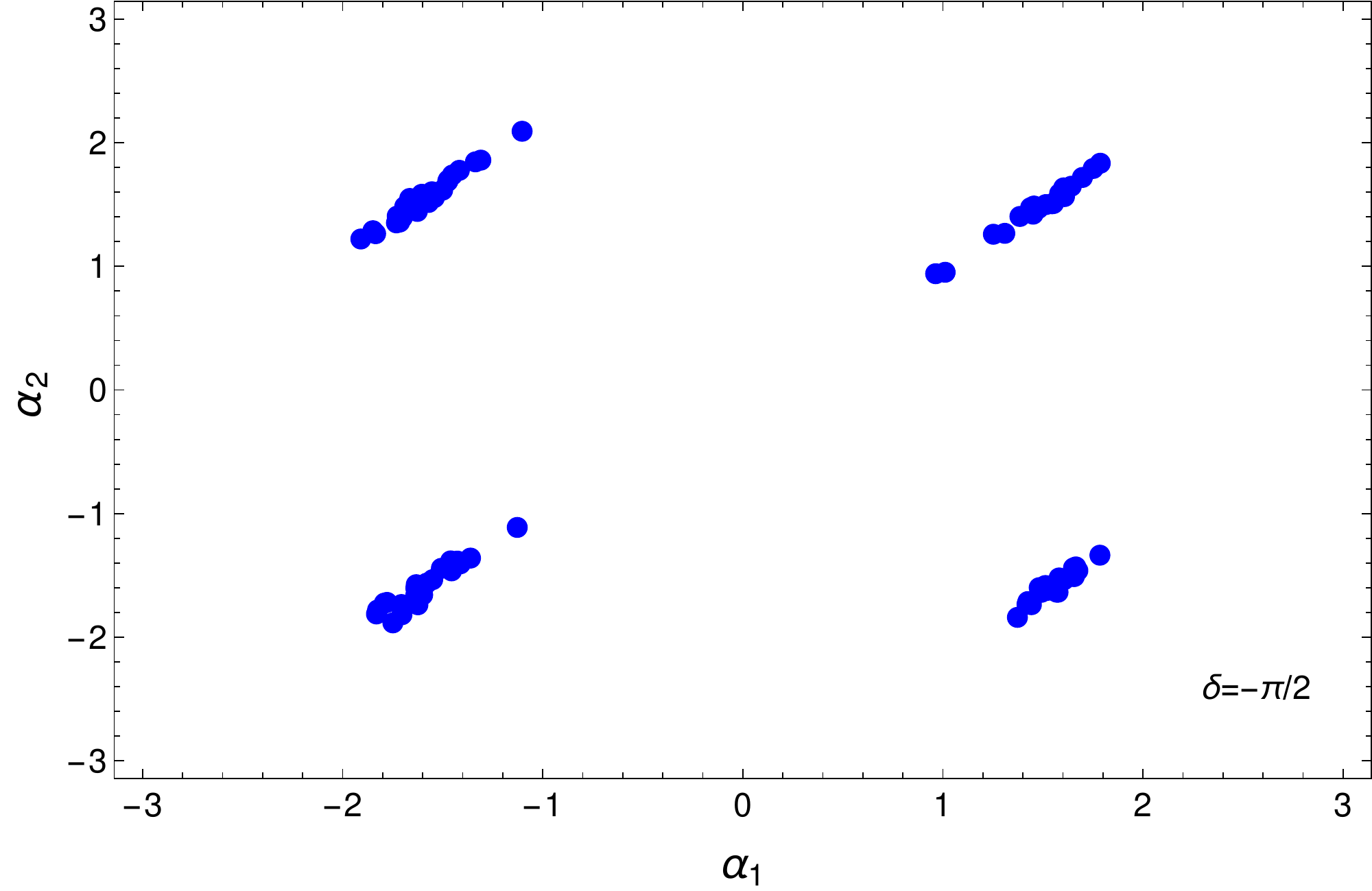}}}
	 	 \qquad
{{\includegraphics[width=70mm]{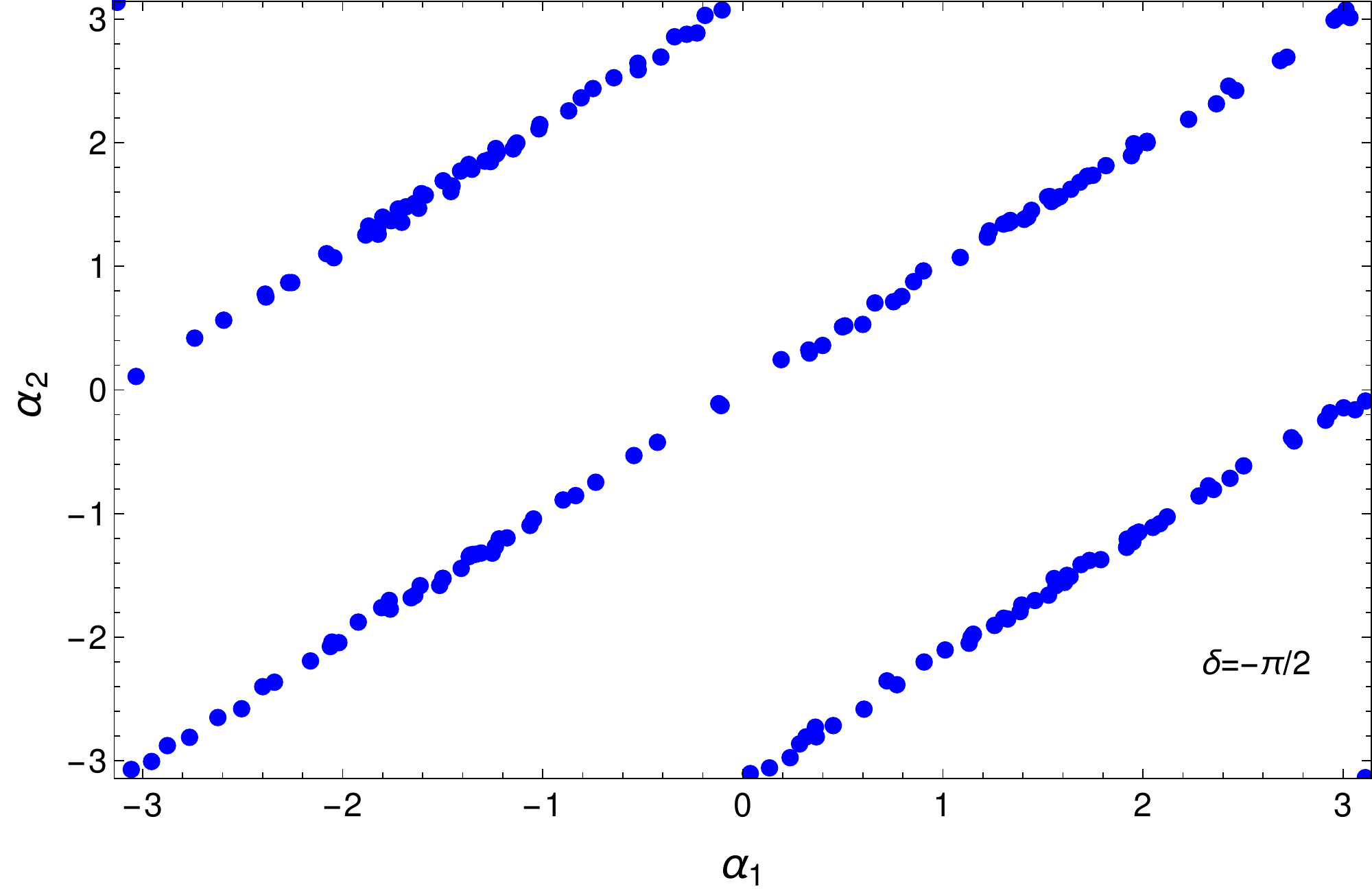}}}
     \caption{\footnotesize{ The allowed values of $\alpha_{1}$ and $\alpha_{2}$, when $|m_{ee}|$ attains its minimum in the ranges: a). $m_{\rm min}=(10^{-2}-1)$ eV, b). $m_{\rm min}=(10^{-3}-10^{-2})$ eV, c). $m_{\rm min}=(10^{-4}-10^{-3})$ eV, d). $m_{\rm min}=(10^{-6}-10^{-4})$ eV. The hierarchy is assumed to be NH. Dirac phase $\delta$ is assumed to be $-\pi/2$.}}
	\label{a1_a2_NH}
\end{figure}

\begin{figure}
	\centering
{{\includegraphics[width=70mm]{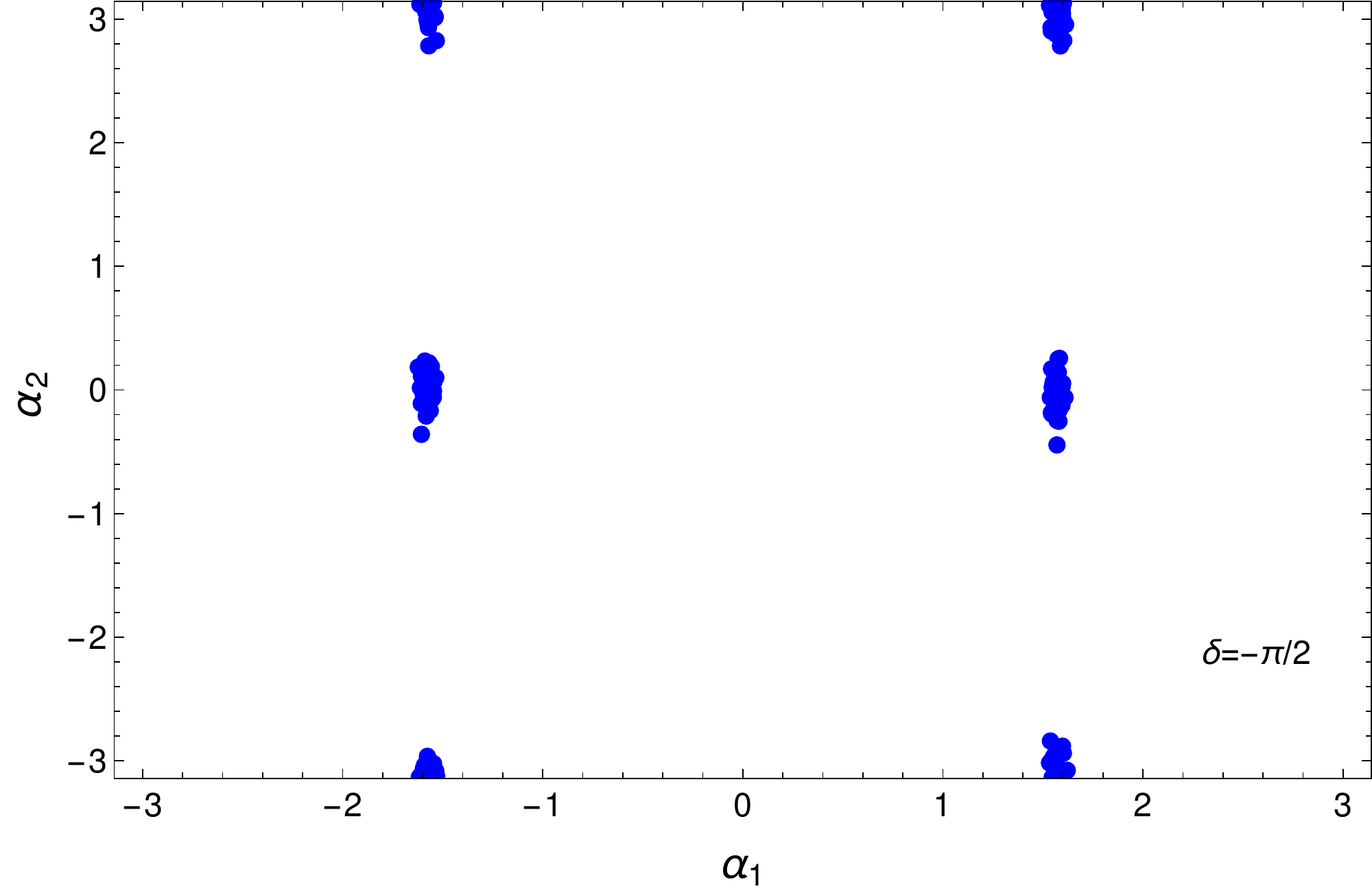}}}
	 	     \qquad
{{\includegraphics[width=70mm]{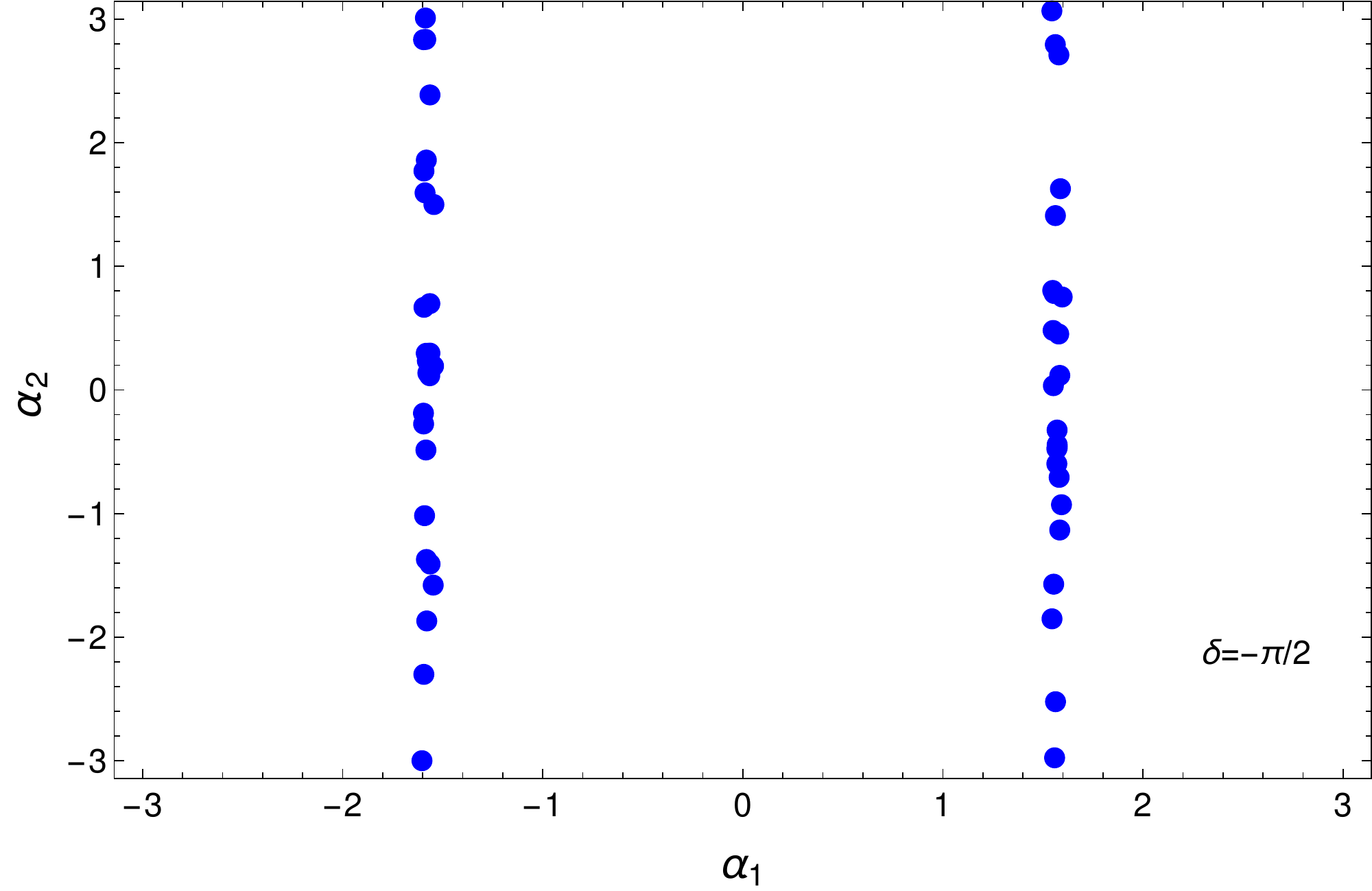}}}
     \caption{\footnotesize{The allowed values of $\alpha_{1}$ and $\alpha_{2}$, when $|m_{ee}|$ attains its minimum in the ranges: a). $m_{\rm min}=(10^{-2}-1)$ eV. b). $m_{\rm min}=(10^{-6}-10^{-2})$ eV. The hierarchy is assumed to be IH. Dirac phase $\delta$ is assumed to be $-\pi/2$.}}
	\label{a1_a2_IH}
\end{figure}

\begin{figure}
	\centering
{{\includegraphics[width=70mm]{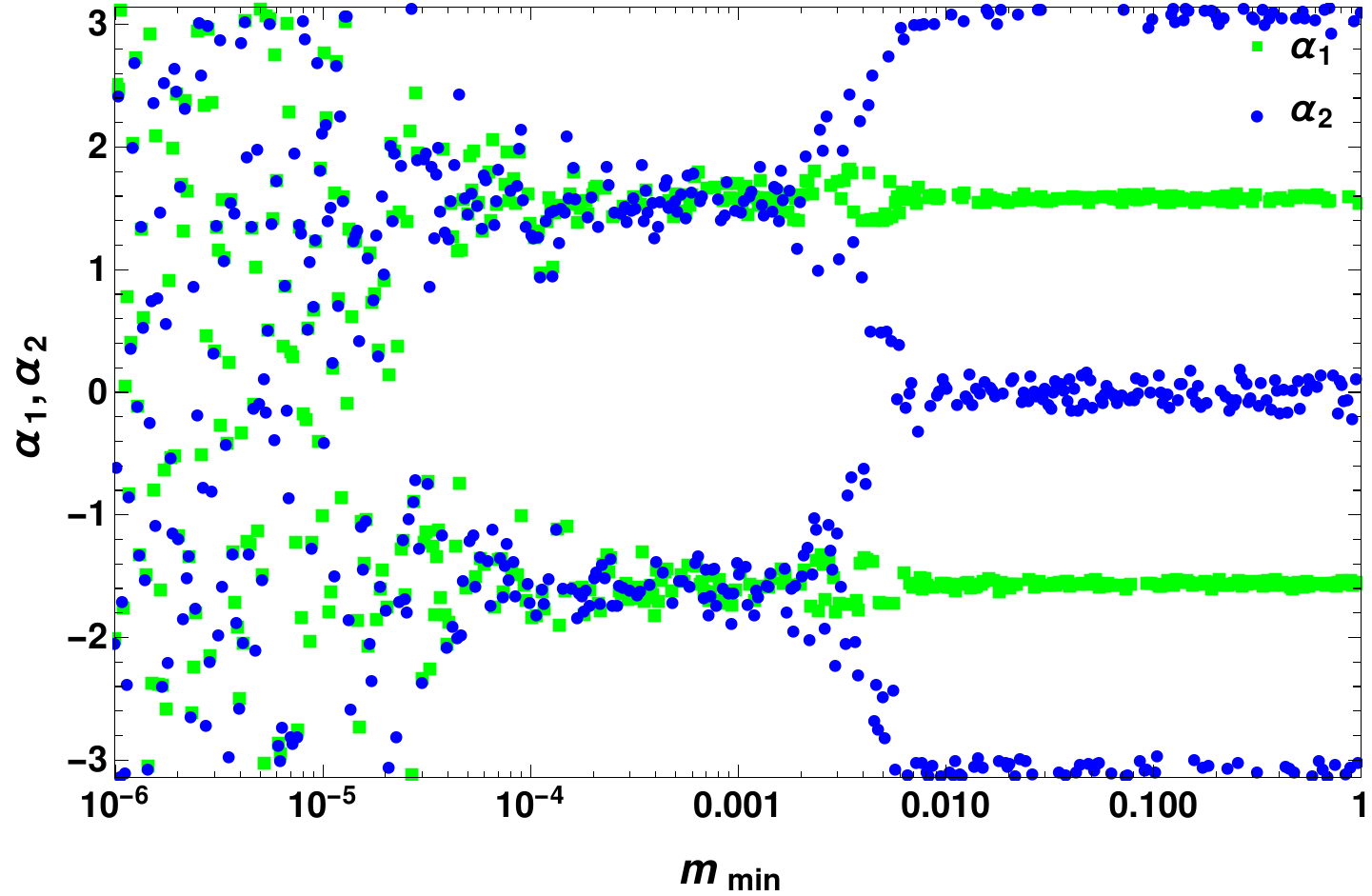}}}
	     \qquad
{{\includegraphics[width=70mm]{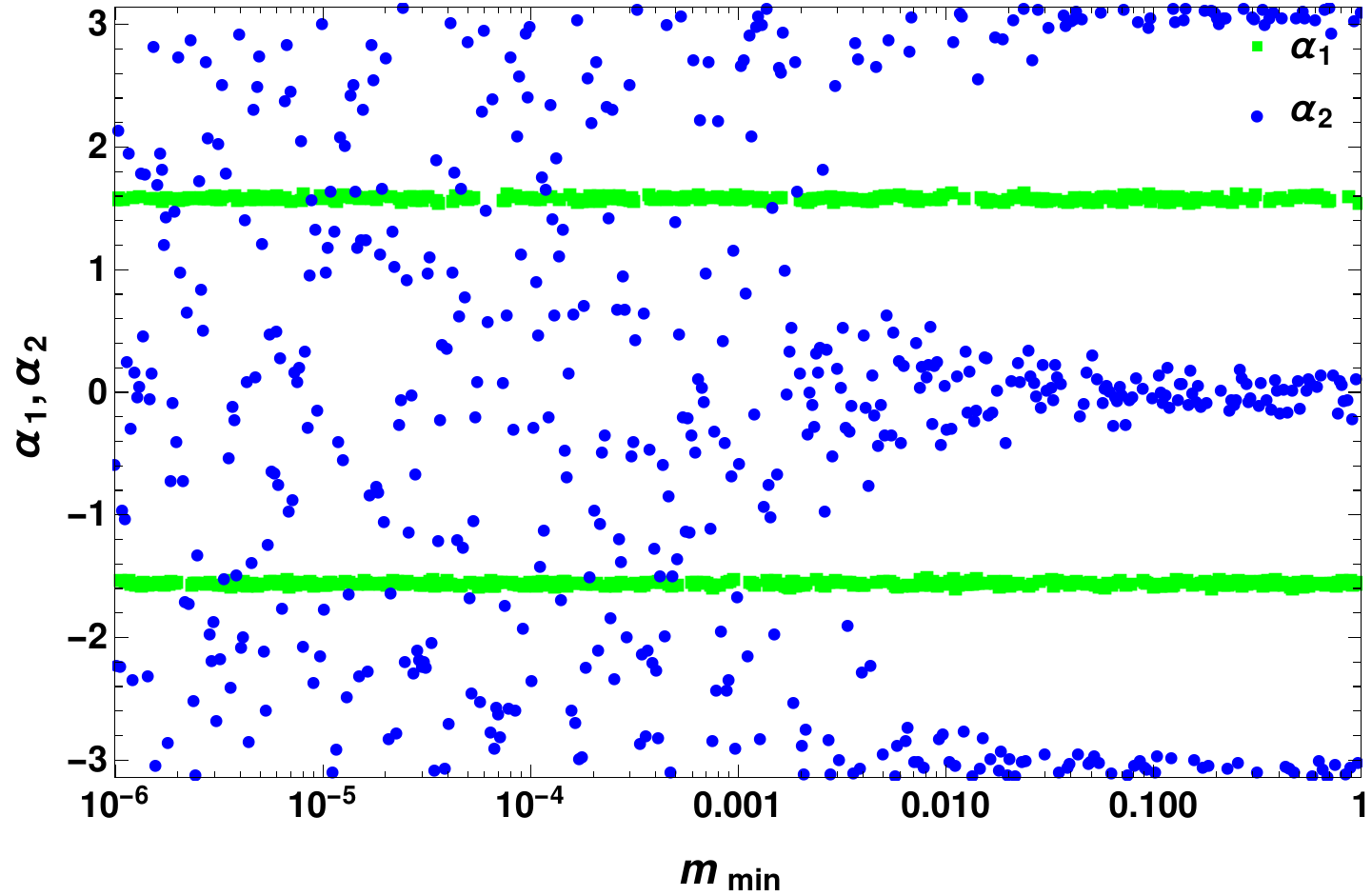}}}
   \caption{\footnotesize{Values of Majorana phases obtained by minimising $|m_{ee}|$: $\alpha_{1}$ (Green) and $\alpha_{2}$ (Blue) versus $m_{\rm min}$ for a). Normal hierarchy, b). Inverted hierarchy. Dirac phase $\delta$ is assumed to be $-\pi/2$.}}
	\label{mmin_a1a2}
\end{figure}

The four panels of Fig.\,\ref{a1_a2_NH} show the values of $\alpha_{1}$ and $\alpha_{2}$ which minimise $|m_{ee}|$ for the four sub-ranges of $m_{\rm min}$ in the case of NH. The two panels of Fig.\,\ref{a1_a2_IH} shows similar results in the case of IH. The relation between $\alpha_{1}$ and $\alpha_{2}$, in all the six panels match those expected from qualitative discussion. The values of Majorana phases, $\alpha_{1}$ and $\alpha_{2}$, obtained by minimising $|m_{ee}|$ are plotted in Fig.\,\ref{mmin_a1a2} as a function of $m_{\rm min}$. We note that the values of $\alpha_{1}$ and $\alpha_{2}$ in Fig.\,\ref{mmin_a1a2} are consistent with those in Fig.\,\ref{a1_a2_NH} (NH) and Fig.\,\ref{a1_a2_IH} (IH). Had we chosen a value of $\delta$ other than $-\pi/2$, the value of $\alpha_{1}$ is unaffected and the value of $\alpha_{2}$ would be shifted by $\delta$. 

\section{Flavoured CP-asymmetry with three right handed neutrinos}\label{CP-asy_3RHN}
In the previous section we derived the constraints on $\alpha_{1}$ and $\alpha_{2}$ as a function of $m_{\rm min}$ by minimizing $|m_{ee}|$. We utilize these values of the phases in the present section to compute the CP-asymmetry paramenter $\epsilon_{1}^{l}$ of leptogenesis. In Type-I seesaw mechanism, the SM is extended by the inclusion of three right handed neutrinos, which have no gauge charges. These neutrinos can have bare Majorana masses. They can also couple to left handed lepton doublet and the Higgs doublet through yukawa couplings. These couplings give rise to a Dirac mass matrix of the neutrinos on spontaneous symmetry breaking. In this extended model, the leptonic mass terms are
\begin{eqnarray}
\mathcal{L_{\rm mass}} &=& -\left( {1\over 2}\overline{(N_{iR})^c}(M_R)_{ij}N_{jR}+ \frac{v}{\sqrt{2}}\,
\overline{\ell_{Li}}
 (Y_e)_{ij}\ell_{Rj}+ \frac{v}{\sqrt{2}}\, \overline{\ell_{Li}} (Y_\nu)_{ij} 
N_{jR}+h.c.\right)\,,
\label{lagrangian}
\end{eqnarray}
where $v$ is the vacuum expectation value of the Higgs. In Eq.\,\ref{lagrangian} $i, j$ run from 1 to 3,  $\ell_{Li}$ represents the ${\rm SU(2)}_L$ doublets, $\ell_{Ri}$ and $N_{jR}$ are right handed charged lepton and neutrino fields respectively. The seesaw mechanism leads to a light neutrino mass matrix, $m_{\nu}=-m_{D}^{T} M_{R}^{-1} m_{D}$, where $m_{D}=Y_{\nu} v/\sqrt{2}$ is the Dirac mass matrix and $M_{R}$ is the mass matrix of right handed neutrinos. Without loss of generality we consider $M_R$ 
to be diagonal and in this basis $m_D$ contains the rest of the physical parameters that appear in $m_\nu$. 

The Majorana mass of the heavy neutrinos $N_i$ ($i=1,2,3)$ can give rise to lepton number violation. Therefore, the CP-violating 
out-of-equilibrium decay of $N_i$ to $\ell \, H$ and ${\bar \ell} H^{\dagger}$ in the early Universe can give rise to a net 
lepton asymmetry. This lepton asymmetry is then converted to an observed baryon asymmetry via electroweak sphalerons. We assume that 
the masses of the right handed neutrinos have the pattern $M_{1}\ll M_{2} \ll M_{3}$, so that the lepton asymmetry arises purely due to the decay of the lightest right handed neutrino.


For a given flavour $l$, the neutrino yukawa coupling matrix can be written in Casas-Ibarra parameterization \cite{Casas:2001sr} as,
\be 
(Y_{\nu})_{i\,l} =  \frac{1}{v}\sqrt{M_i}R_{ij}\sqrt{m_j} U_{l\, j}^{*}\,.
\label{casas}
\ee  
In Eq.\,\ref{casas} $m_j$, $M_{i}, (j,i=1,2,3)$ are mass eigenvalues of the light and heavy Majorana neutrinos respectively and $U$ is the PMNS matrix. The matrix $R$ in Eq.\,\ref{casas} is a complex orthogonal matrix. It is parameterized in terms of three complex angles $z_{i}$ $(i=1,2,3)$ as 
\begin{equation}
			R=
  			\begin{pmatrix}
    			\cos z_{2} \cos z_{3} & \cos z_{2} \sin z_{3} & \sin z_{2} \\
    			-\cos z_{3} \sin z_{1} \sin z_{2}-\cos z_{1} \sin z_{3} & \cos z_{1} \cos z_{3}-\sin z_{1} \sin z_{2} \sin z_{3} & \cos z_{2} \sin z_{1} \\
    			-\cos z_{1} \cos z_{3} \sin z_{2}+\sin z_{1} \sin z_{3} & -\cos z_{3} \sin z_{1}-\cos z_{1}\sin z_{2} \sin z_{3} & \cos z_{1} \cos z_{2} 
  			\end{pmatrix}.
  			\label{R}
\end{equation}
 
The CP-asymmetry generated in a particular flavour $l$ $(l=e,\mu,\tau)$, is given by 
\be
\epsilon_{1}^{l}= -\frac{3M_1}{16 \pi v^2}\,\frac{{\rm Im} \left(\sum\limits_{j k}m_j^{1/2}m_{k}^{3/2}\,U^*_{lj}U_{lk}R_{1j}R_{1k}\right)}{\sum\limits_{j} m_{j} |R_{1j}|^2},
\label{cp_asy_a}
\ee
where $m_{j}$ and $m_{k}$ are appropriate light neutrino mass eigen value. 

In terms of Casas-Ibarra parameterization, Eq.\,\ref{cp_asy_a} can be written as\,\cite{Pascoli:2006ci}
	\begin{eqnarray}
\epsilon_{1}^{l}&=& -\frac{3 M_{1}}{16 \pi v^{2}}\,\, \frac{1}{m_{1}|R_{11}|^{2}+m_{2}|R_{12}|^{2}+m_{3}|R_{13}|^{2}} \nonumber\\
&\times & \lbrace m_{1}^{2}\,{\rm Im}[U_{l1}^{*}U_{l1}R_{11}R_{11}]+m_{1}^{1/2}m_{2}^{3/2}\,{\rm Im}[U_{l1}^{*}U_{l2}R_{11}R_{12}]+m_{1}^{1/2}m_{3}^{3/2} \,{\rm Im}[U_{l1}^{*}U_{l3}R_{11}R_{13}] \nonumber\\
&& + m_{2}^{1/2}m_{1}^{3/2} \,{\rm Im}[U_{l2}^{*}U_{l1}R_{12}R_{11}]+m_{2}^{2} \,{\rm Im}[U_{l2}^{*}U_{l2}R_{12}R_{12}]+m_{2}^{1/2}m_{3}^{3/2} \,{\rm Im}[U_{l2}^{*}U_{l3}R_{12}R_{13}] \nonumber\\
&& + m_{3}^{1/2} m_{1}^{3/2}\,{\rm Im}[U_{l3}^{*}U_{l1}R_{13}R_{11}]+m_{3}^{1/2} m_{2}^{3/2} \,{\rm Im}[U_{l3}^{*}U_{l2}R_{13}R_{12}]+m_{3}^{2} \,{\rm Im}[U_{l3}^{*}U_{l3}R_{13}R_{13}]\rbrace .
	\label{cp_asy_b}
	\end{eqnarray}
We assume that the $\epsilon^{l}_{1}$ arises only through the phases in the PMNS matrix. Thus, we set the phases in $R$ to be zero and assume $z_{i}$ to be real. Hence $R$ becomes a real orthogonal matrix. Only the decays of the lightest right handed neutrino creates CP-asymmetry. Therefore only the elements of the first row of $R$ enter in the expression $\epsilon_{1}^{l}$. In our parameterization of $R$, these elements depend only on $\sin z_{2}$ and $\sin z_{3}$, {\it i.e.,} $\epsilon_{1}^{l}$ is independent of $\sin z_{1}$. For simplicity, we assume $\sin z_{2}=\sin z_{3}=\sin z$. It must be noted that the total CP-asymmetry, $\epsilon_{1}=\epsilon^{e}_{1}+\epsilon^{\mu}_{1}+\epsilon^{\tau}_{1}=0$ when the matrix $R$ is real. Under the assumptions we made, the expression in Eq.\,\ref{cp_asy_b} simplifies to 
	\begin{eqnarray}
	\epsilon_{1}^{l}&=& -\frac{3 M_{1}}{16 \pi v^{2}} \frac{1}{m_{1}|R_{11}|^{2}+m_{2}|R_{12}|^{2}+m_{3}|R_{13}|^{2}} \nonumber\\
&\times &  \lbrace\sqrt{m_{1}m_{2}} \left(m_{2}-m_{1} \right){\rm Im}[U_{l1}^{*}U_{l2}] \,R_{11}R_{12} \nonumber\\
&& + \sqrt{m_{1}m_{3}}\left(m_{3}-m_{1} \right){\rm Im}[U_{l1}^{*}U_{l3}] \,R_{11}R_{13} \nonumber\\
&& + \sqrt{m_{2}m_{3}} \left(m_{3}-m_{2} \right){\rm Im}[U_{l2}^{*}U_{l3}] \,R_{12}R_{13} \rbrace .
	\label{cp_asy_d}
	\end{eqnarray}	
	
\begin{figure}
{{\includegraphics[width=65mm]{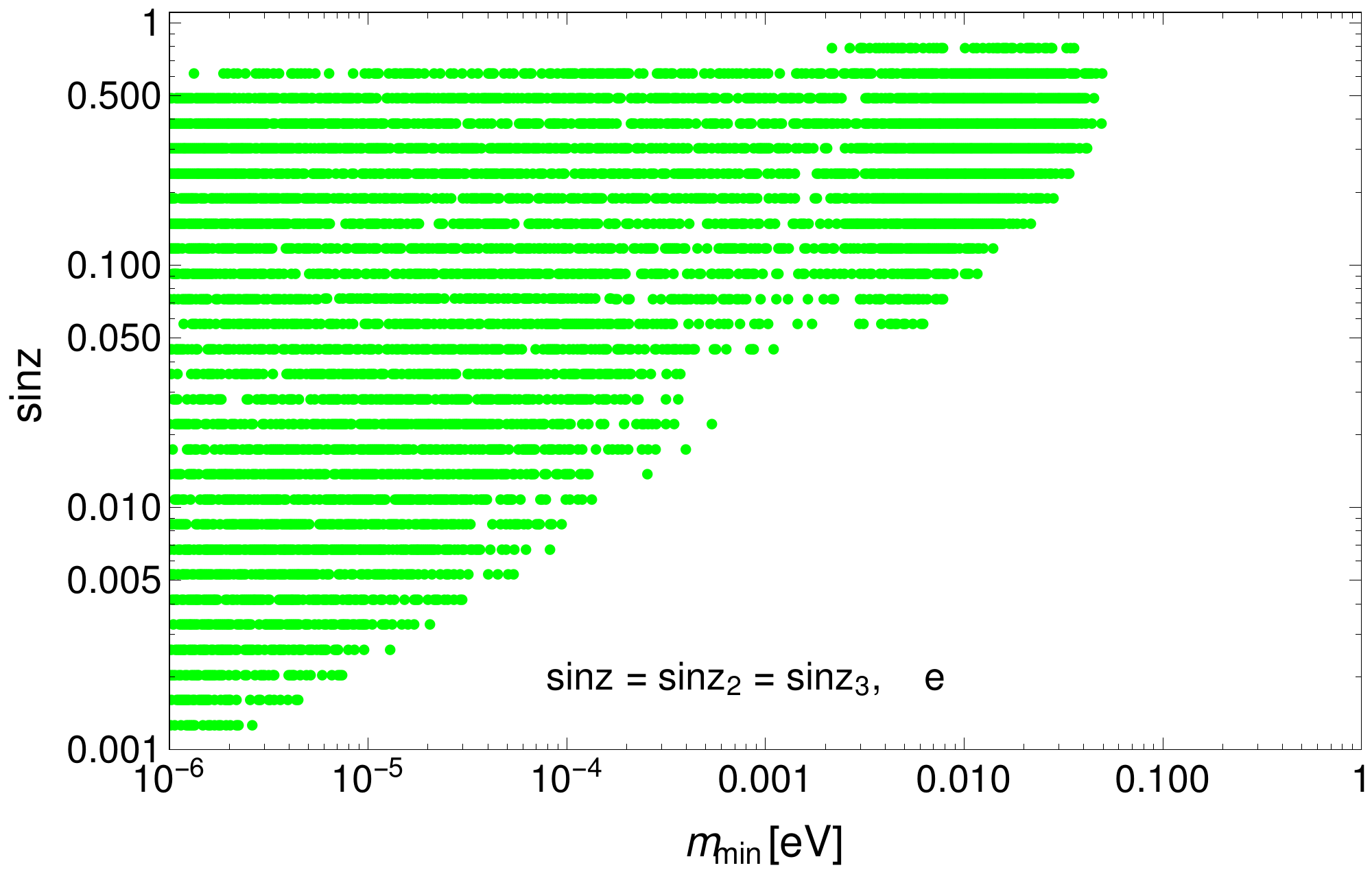}}}
{{\includegraphics[width=65mm]{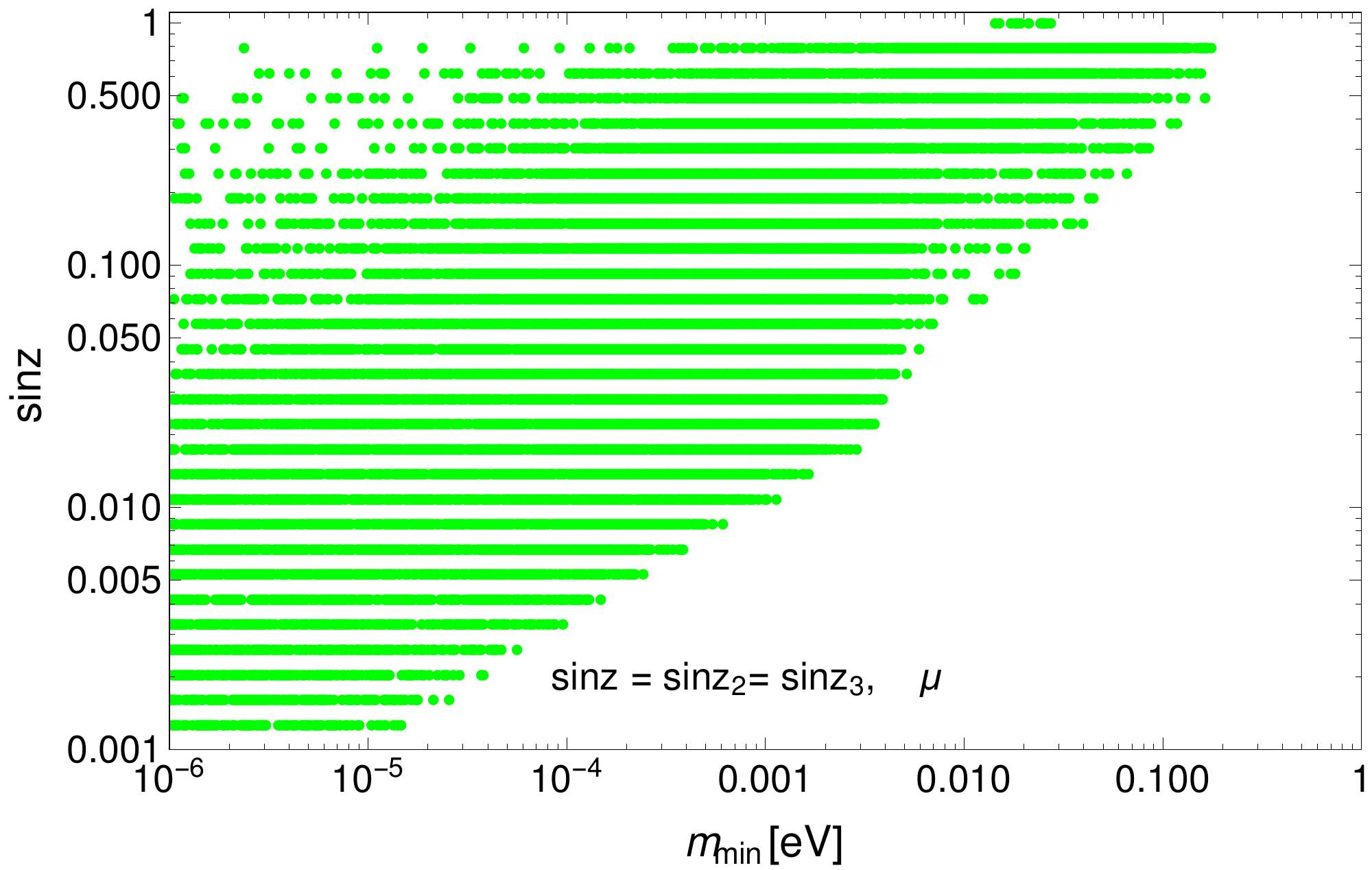}}}
   \qquad
{{\includegraphics[width=65mm]{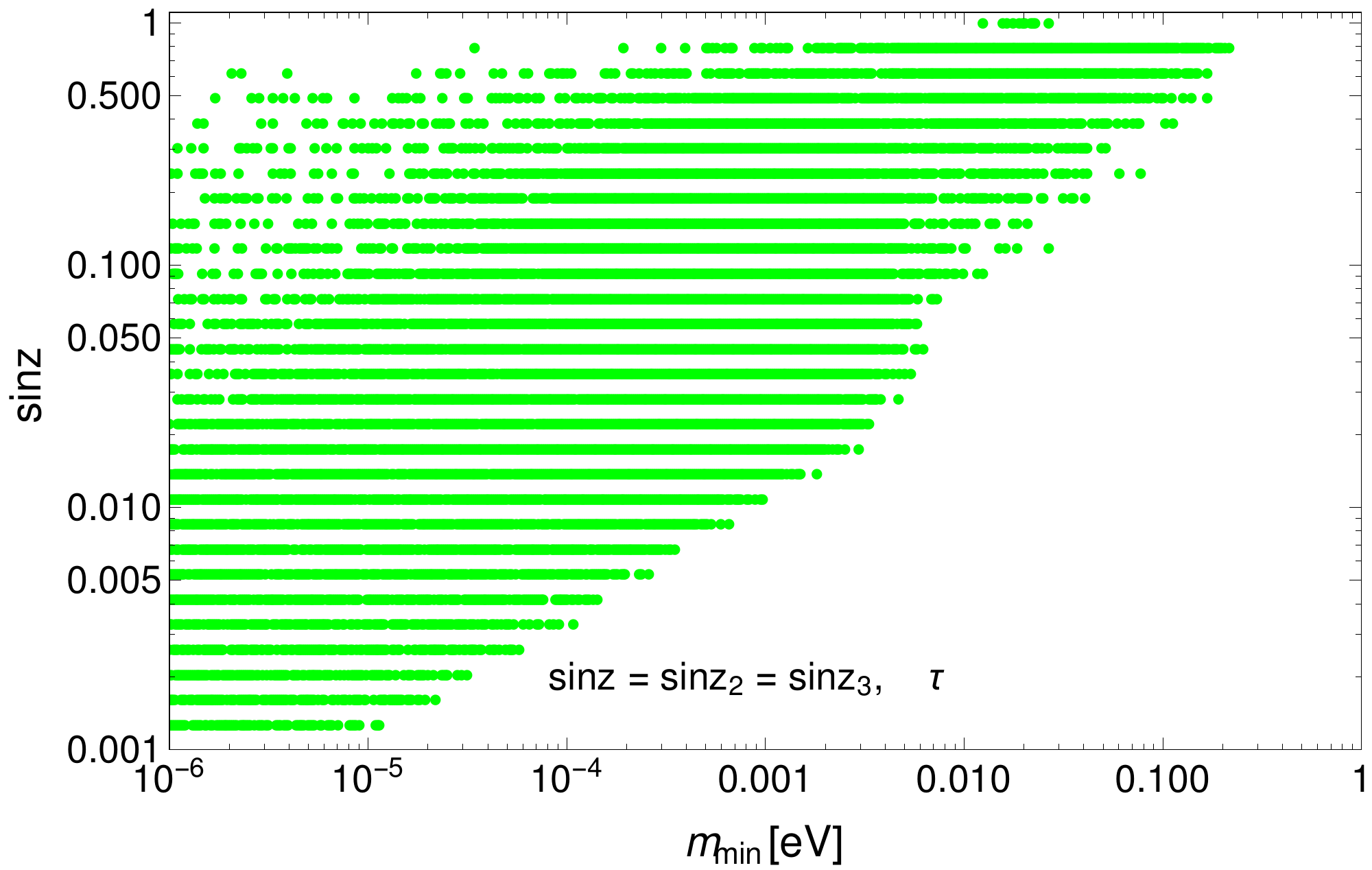}}}
    \caption{\footnotesize{Allowed points in the plane of $\sin z$ versus $m_{\rm min}$ when $|m_{ee}|$ is minimised,
which give rise to $\epsilon_{1} > 10^{-8}$ $\epsilon^{l}, l=e ({\rm Fig}. a),\mu ({\rm Fig}. b),\tau ({\rm Fig}. c)$ . We assumed the
hierarchy to be NH and used the value $M_{1}=10^{10}$ GeV (Green) . We also set $\sin z_{2}$=$\sin z_{3}$=$\sin z$.} }
	\label{mmin_vs_resinz_NH}
\end{figure}

\begin{figure}
{{\includegraphics[width=65mm]{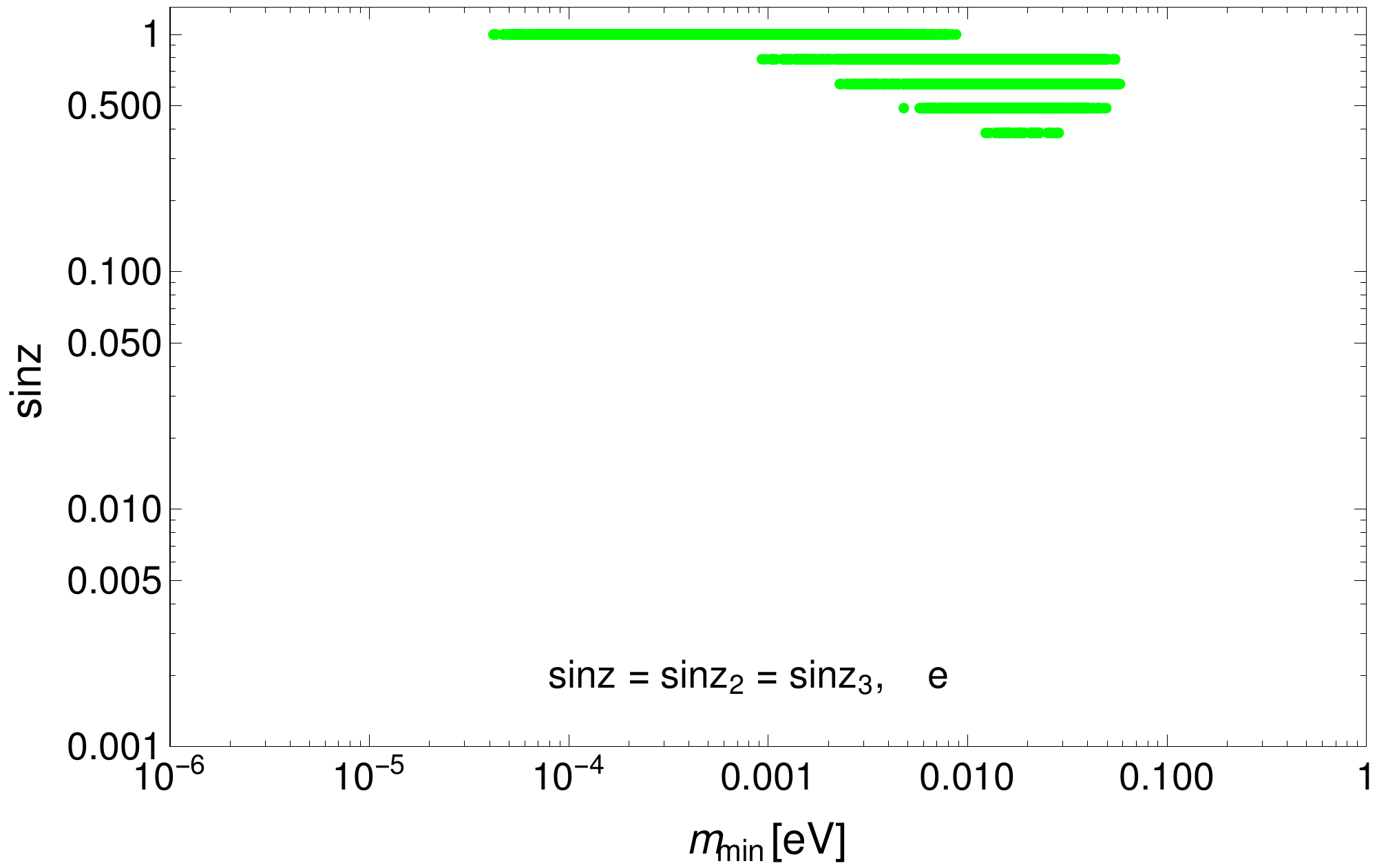}}}
{{\includegraphics[width=65mm]{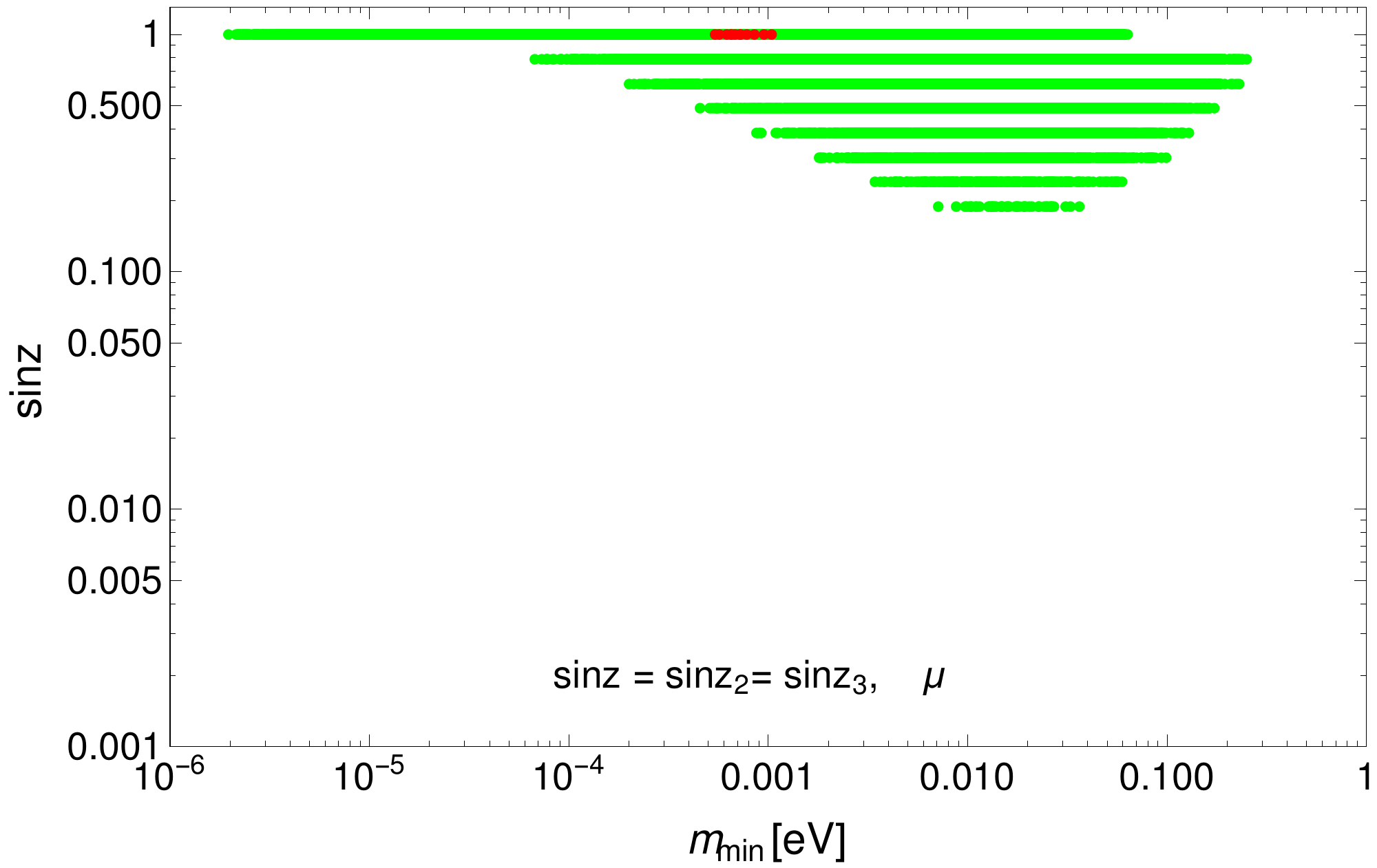}}}
    \qquad
{{\includegraphics[width=65mm]{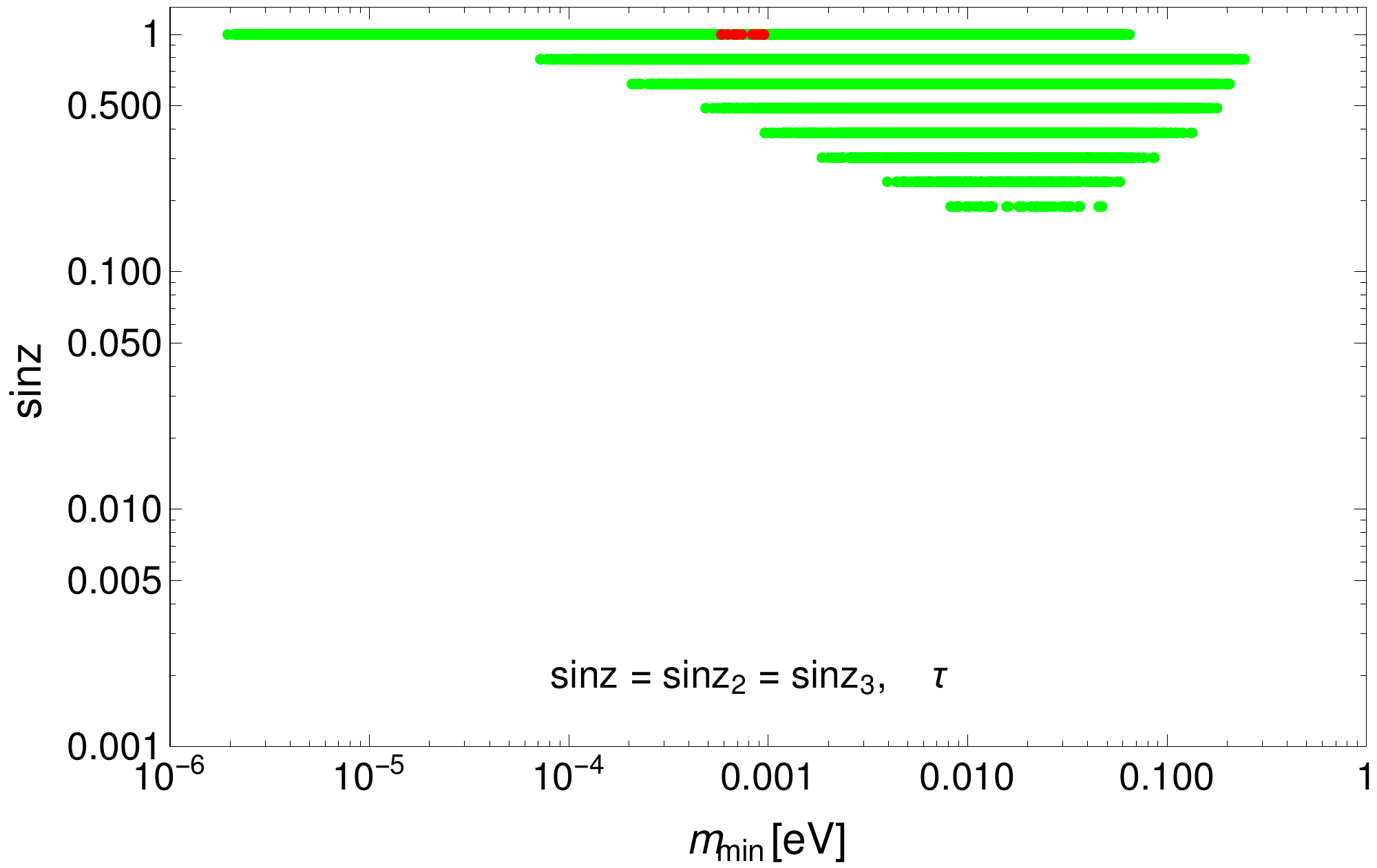}}}
     \caption{\footnotesize{Allowed points in the plane of $\sin z$ versus $m_{\rm min}$ when $|m_{ee}|$ is 
minimized, which give rise to $\epsilon_{1}^{l} > 10^{-8}$, $l=e ({\rm Fig}. a),\mu ({\rm Fig}. b),\tau ({\rm Fig}. c)$. We assumed the
hierarchy to be IH and used two value of $M_1$: $M_{1}=10^{9}$ GeV (Red) and $M_{1}=10^{10}$ GeV (Green). We set $\sin z_{2}$=$\sin z_{3}$=$\sin z$.}}
     \label{mmin_vs_resinz_IH}
\end{figure}

	This expression contains various low energy parameters, the mass of the lightest right handed neutrino, the Higgs vacuum expectation value and an extra free parameter $\sin z$ which comes from high energy scale. Here we study the dependence of CP-asymmetry on the parameters $m_{\rm min}$ and $\sin z$. In our calculation, the elements of PMNS matrix are computed for $\delta=-\pi/2$ and the values of $\alpha_{1}$ and $\alpha_{2}$, given by the condition of minimizing $|m_{ee}|$. We consider only those values of $m_{\rm min}$ and $\sin z$ for which $\epsilon_{1}\gtrsim 10^{-8}$. We take into account the Davidson-Ibarra bound \cite{Davidson:2002qv} by considering the values of $M_{1} \geq 10^{8}$ GeV. Our results are shown in Fig.\,\ref{mmin_vs_resinz_NH} for NH and and Fig.\,\ref{mmin_vs_resinz_IH} for IH.

We considered three values of $M_{1}=10^{8}, 10^{9}, 10^{10}$ GeV. When $M_{1}=10^{8}$ GeV, all values of $\epsilon_{1}^{l}$ $(l=e, \mu. \tau)$ are less than $10^{-8}$ independent of $m_{\rm min}$ and $\sin z$, both for NH and IH. For $M_{1}=10^{9}$ GeV, the asymmetries $\epsilon_{1}^{\mu}, \, \epsilon_{1}^{\tau}$ become larger than $10^{-8}$ only for IH and that too for a single 
point $\sin z =1$ and $m_{\rm min} \simeq 10^{-3}$ eV. For $M_{1}=10^{10}$ GeV, a large set of values of $\sin z$ and
$m_{\rm min}$ are allowed. For NH, all three asymmetries have values in the range $10^{-8}$ to $10^{-7}$ for the  
$m_{\rm min}$ range $(10^{-6},0.1)$ eV and $\sin z$ range $(10^{-3},1)$. In the case of IH, $\epsilon_{1}^{l}>10^{-8}$ is possible only if $\sin z > 0.2$ and $m_{\rm min}$ in the range $(10^{-4},0.2)$ eV. Hence the condition of minimization of $|m_{ee}|$ requires the lightest heavy neutrino mass $M_{1}$ to be $\gtrsim 10^{10}$ GeV to give rise to adequate leptogenesis. From Eq.\,\ref{cp_asy_d} we see that $\epsilon_{1}^{l} \rightarrow 0$ if the light neutrino masses are almost degenerate. Hence there are no allowed points for $m_{\rm min} \geq 0.2$ eV for both NH and IH. 

\begin{figure}
	\centering
{{\includegraphics[width=70mm]{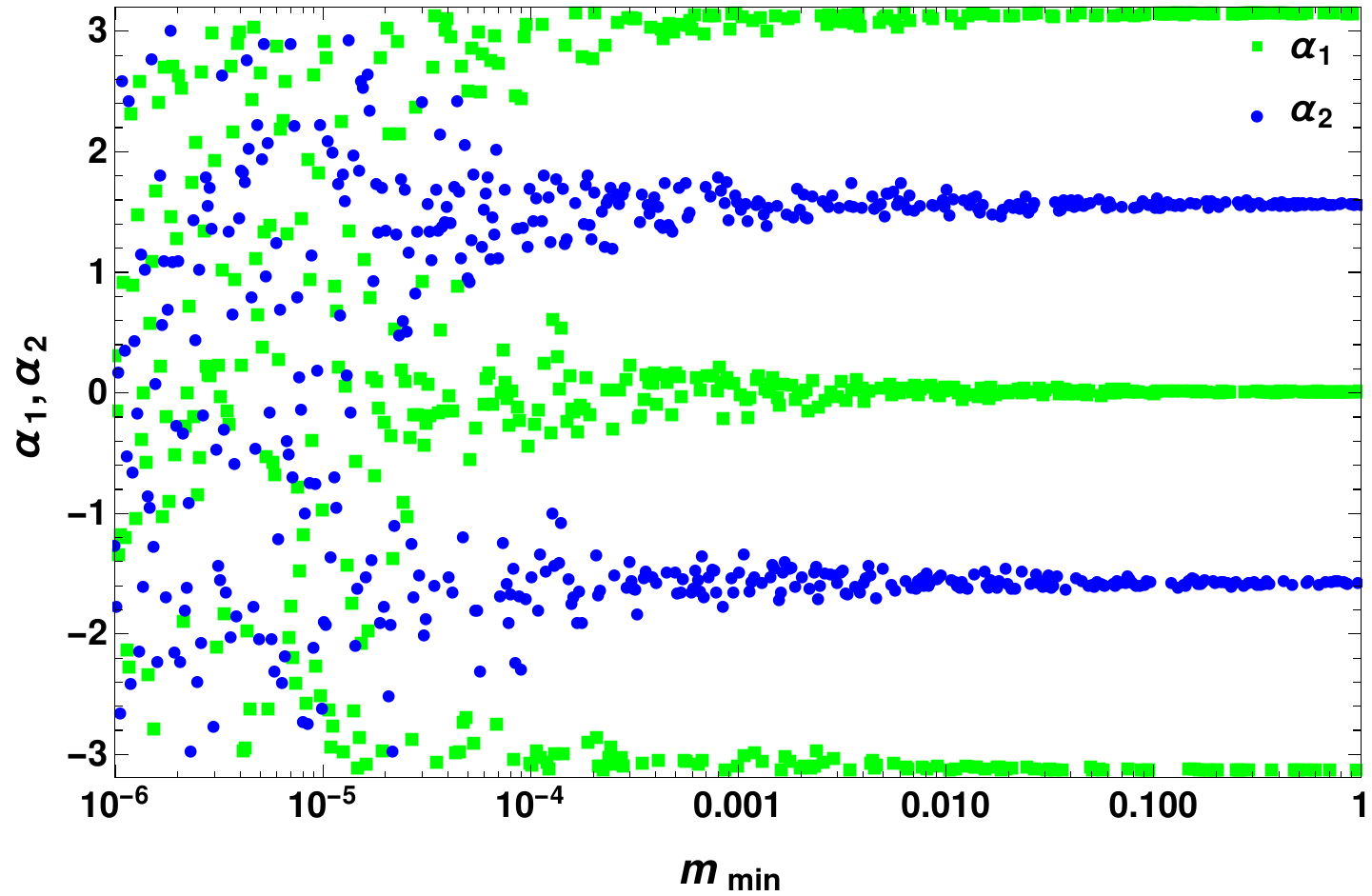}}}
	     \qquad
{{\includegraphics[width=70mm]{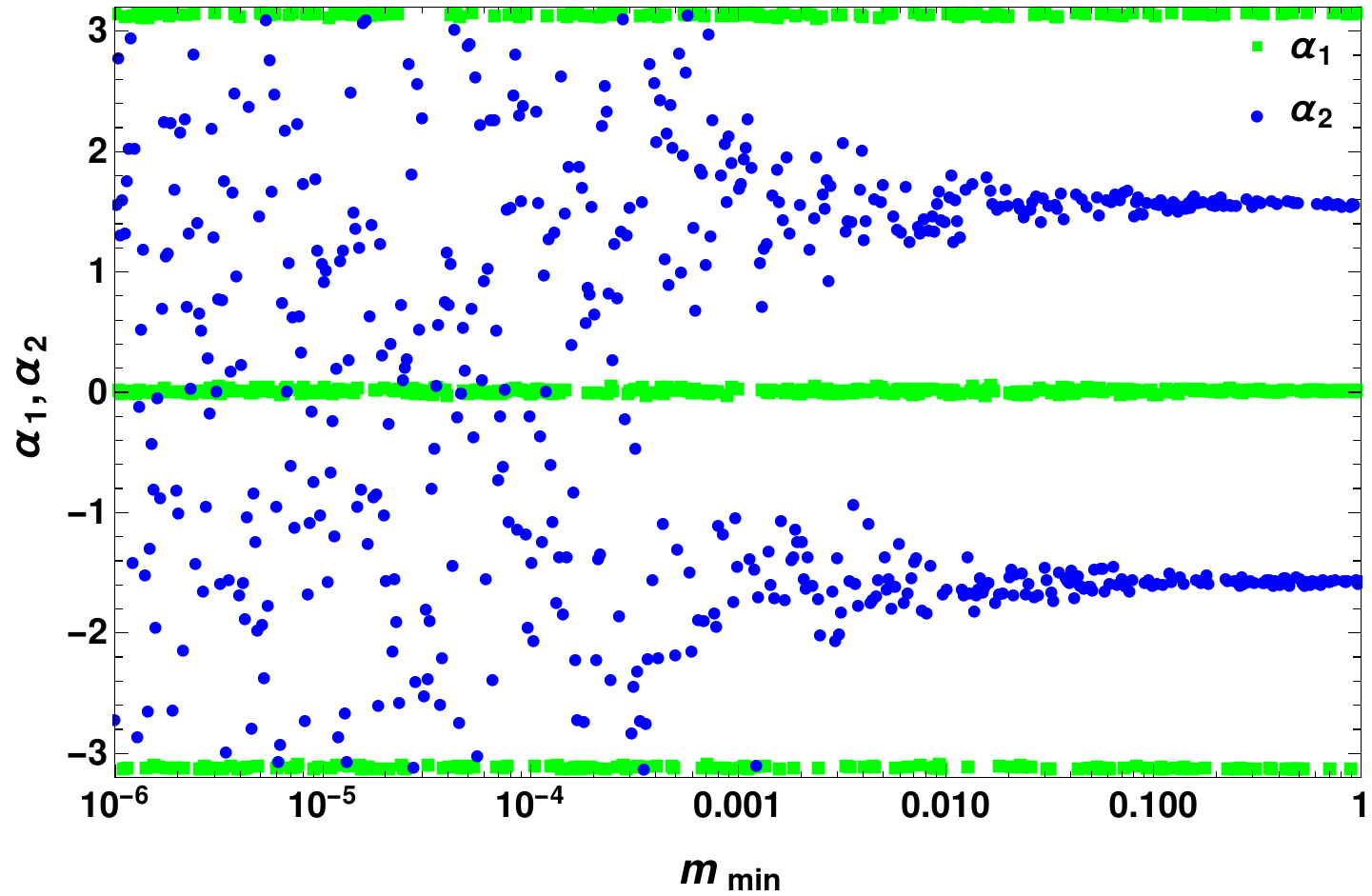}}}
   \caption{\footnotesize{Values of Majorana phases obtained by maximising $|m_{ee}|$: 
$\alpha_{1}$(Green) and $\alpha_{2}$(Blue) versus $m_{\rm min}$: a). Normal hierarchy, b). Inverted hierarchy. 
Dirac phase $\delta$ is assumed to be $-\pi/2$.}}
	\label{mmin_a1a2_max}
\end{figure}

\begin{figure}
{{\includegraphics[width=65mm]{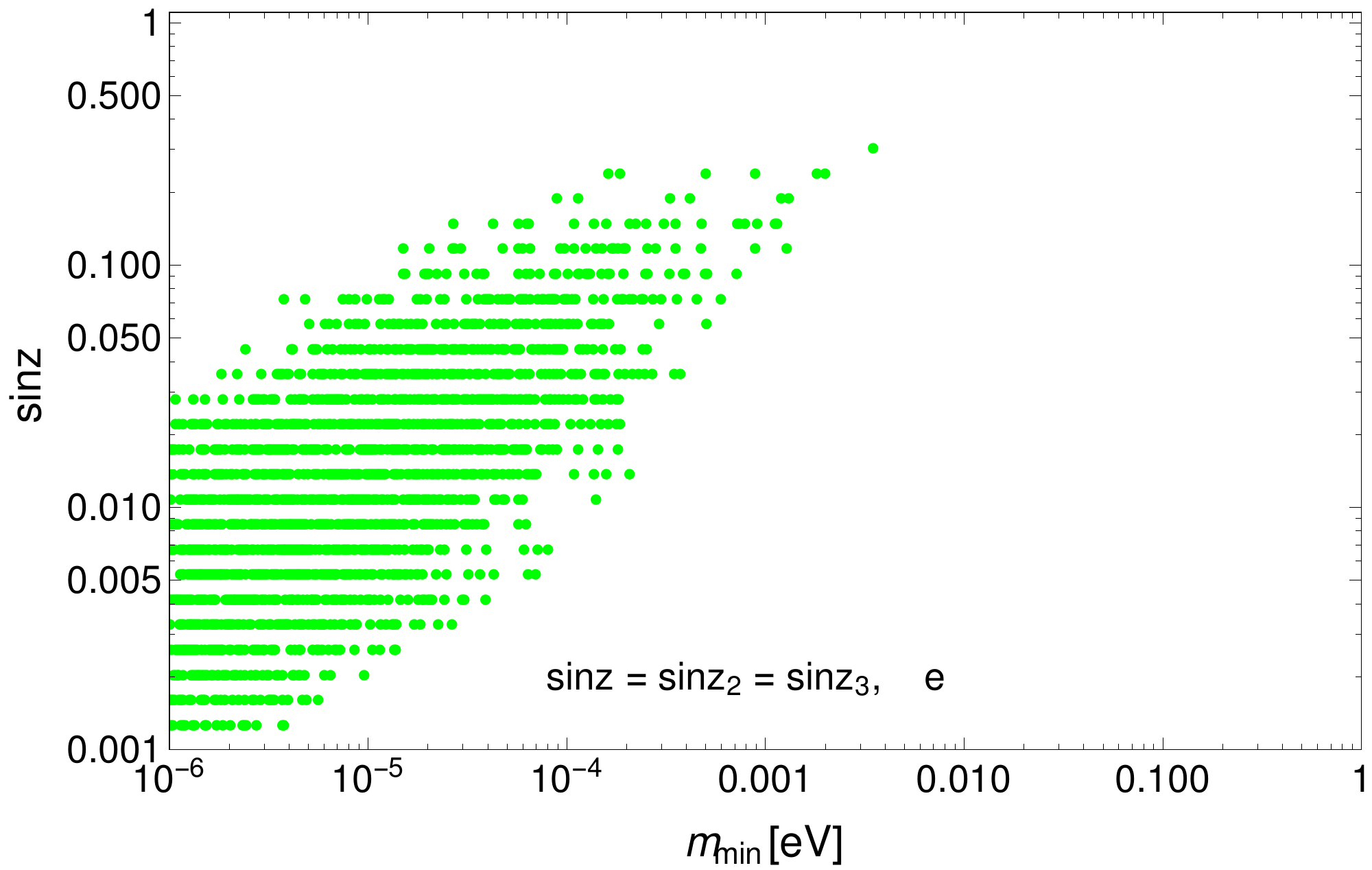}}}
{{\includegraphics[width=65mm]{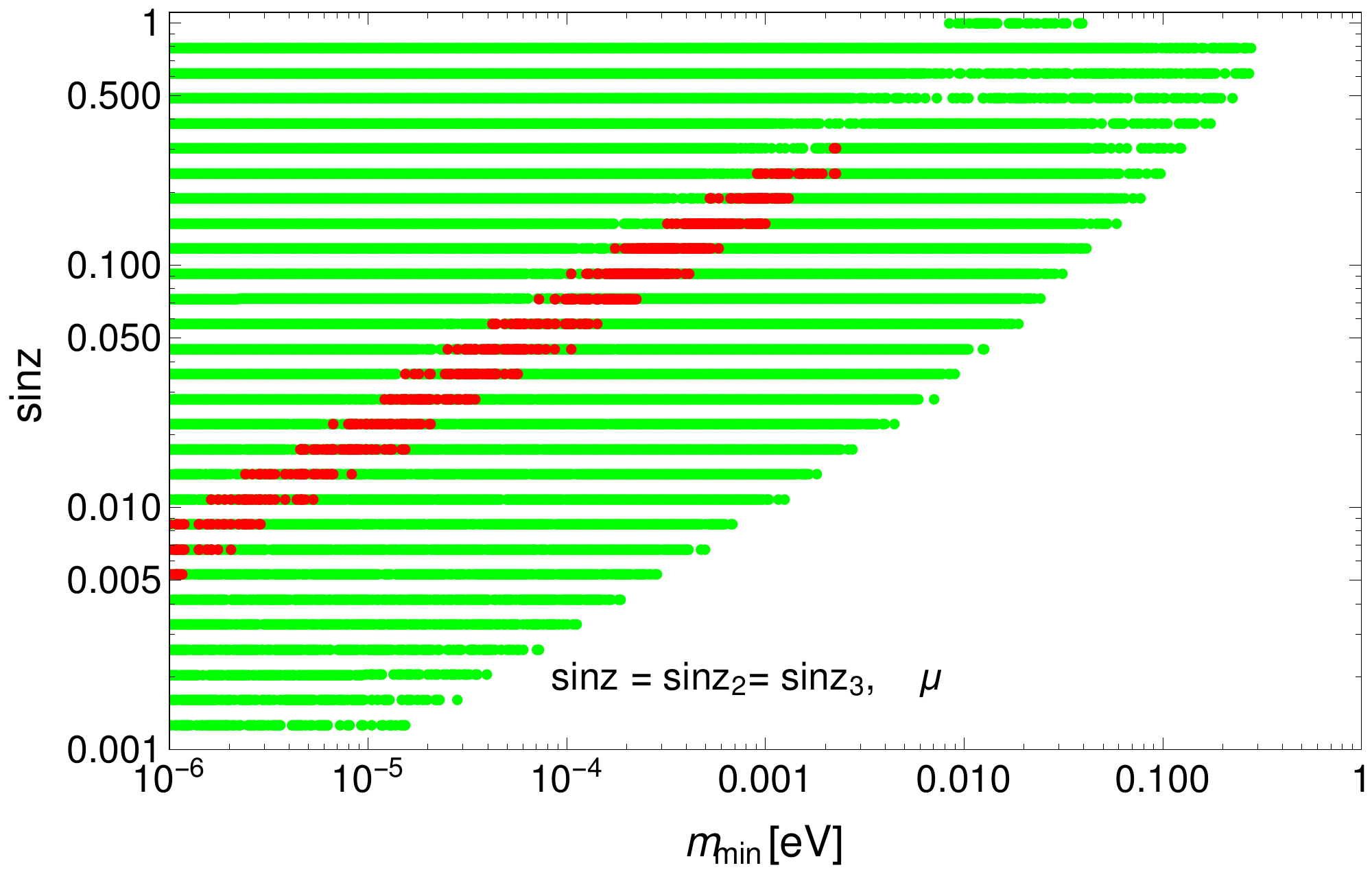}}}
   \qquad
{{\includegraphics[width=65mm]{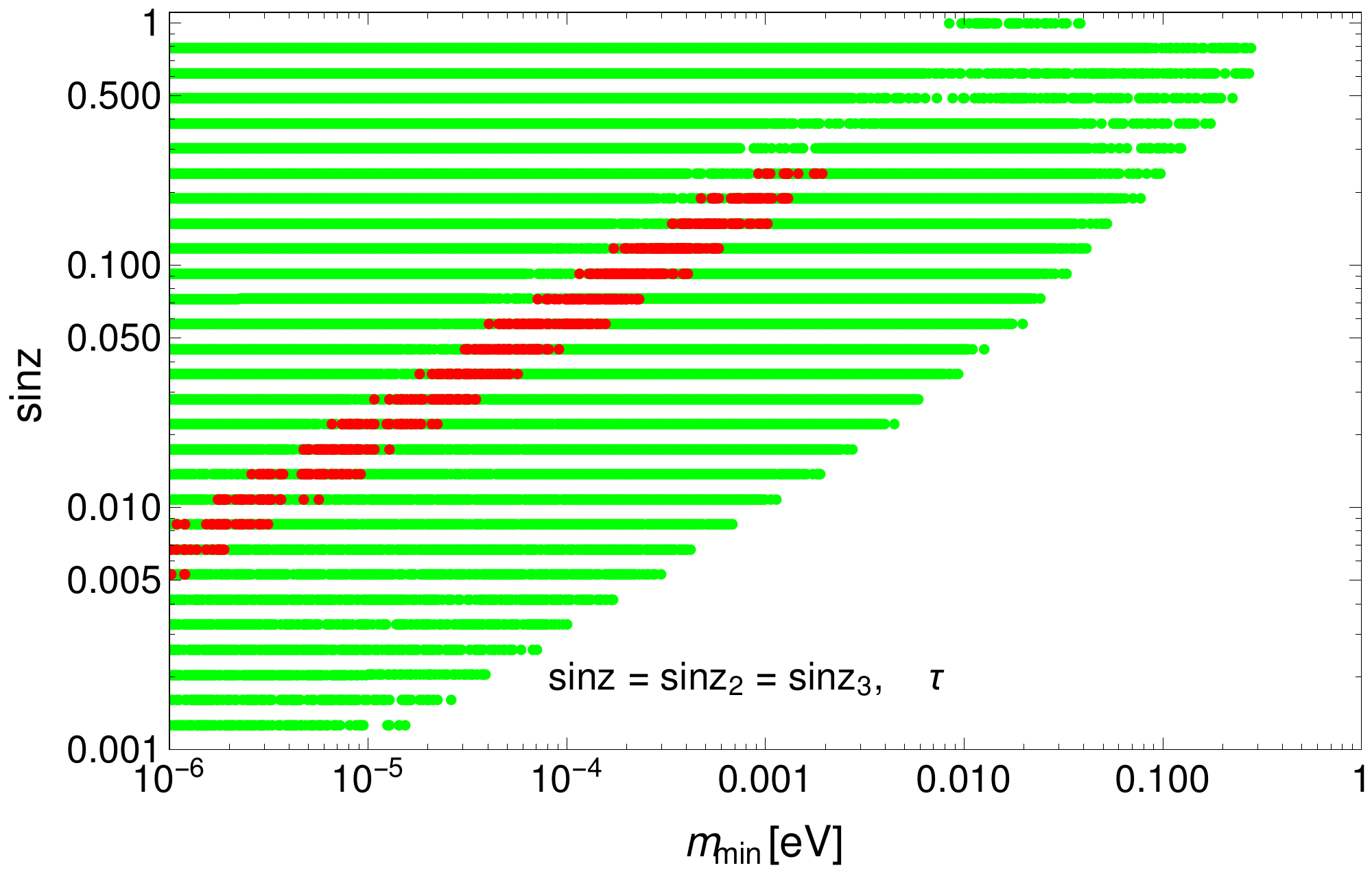}}}
\caption{\footnotesize{Allowed points in the plane of $\sin z$ versus $m_{\rm min}$ when $|m_{ee}|$ is maximised,
which give rise to $\epsilon_{1} > 10^{-8}$ $\epsilon^{l}, l=e ({\rm Fig}. a),\mu ({\rm Fig}. b),\tau ({\rm Fig}. c)$ . We assumed the
hierarchy to be NH and used two values for $M_1$: $M_1 = 10^9$ GeV (Red) and $M_{1}=10^{10}$ GeV (Green) . We also set $\sin z_{2}$=$\sin z_{3}$=$\sin z$.} }
	\label{mmin_vs_resinz_NH_meemax}
\end{figure}

\begin{figure}
{{\includegraphics[width=65mm]{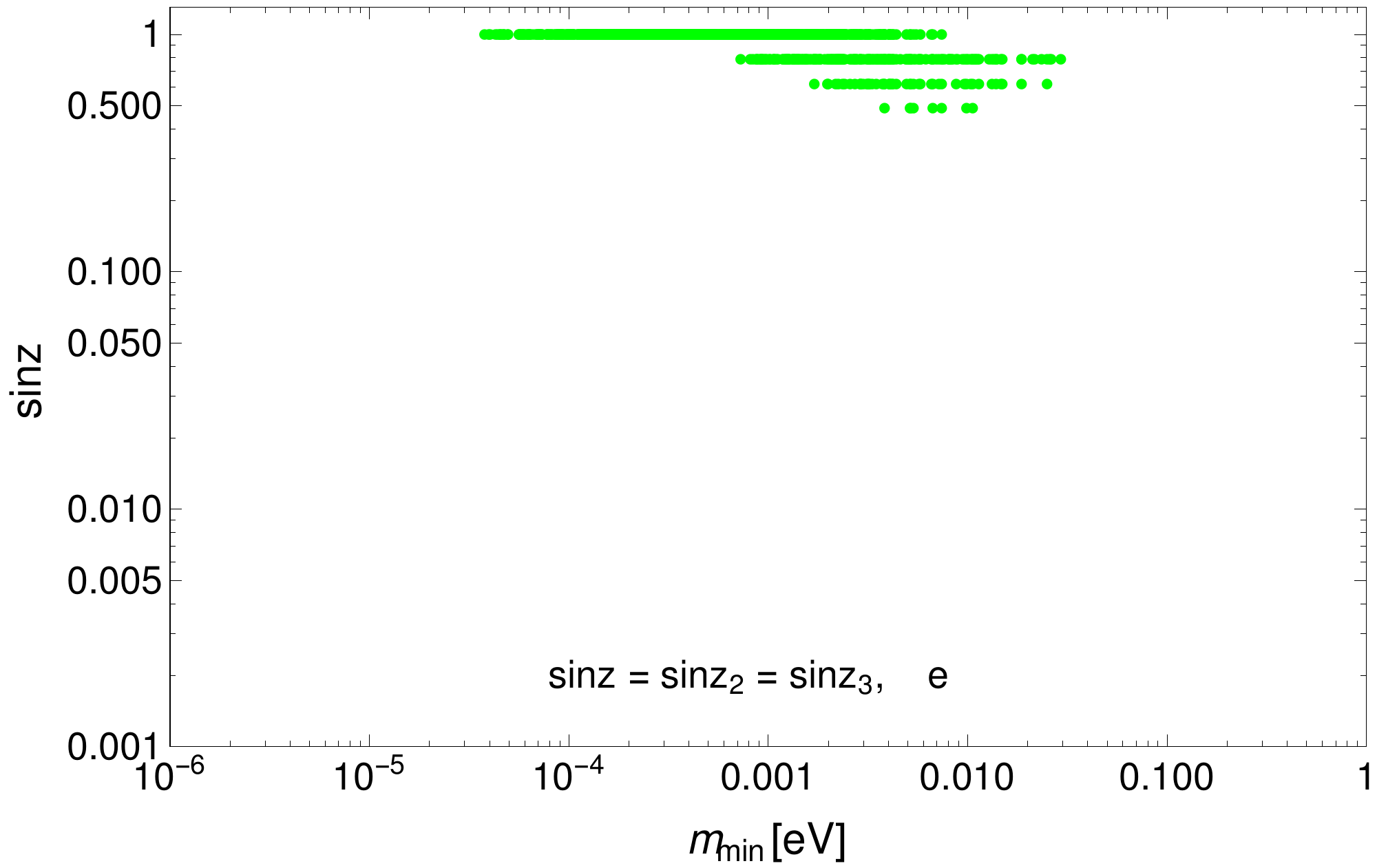}}}
{{\includegraphics[width=65mm]{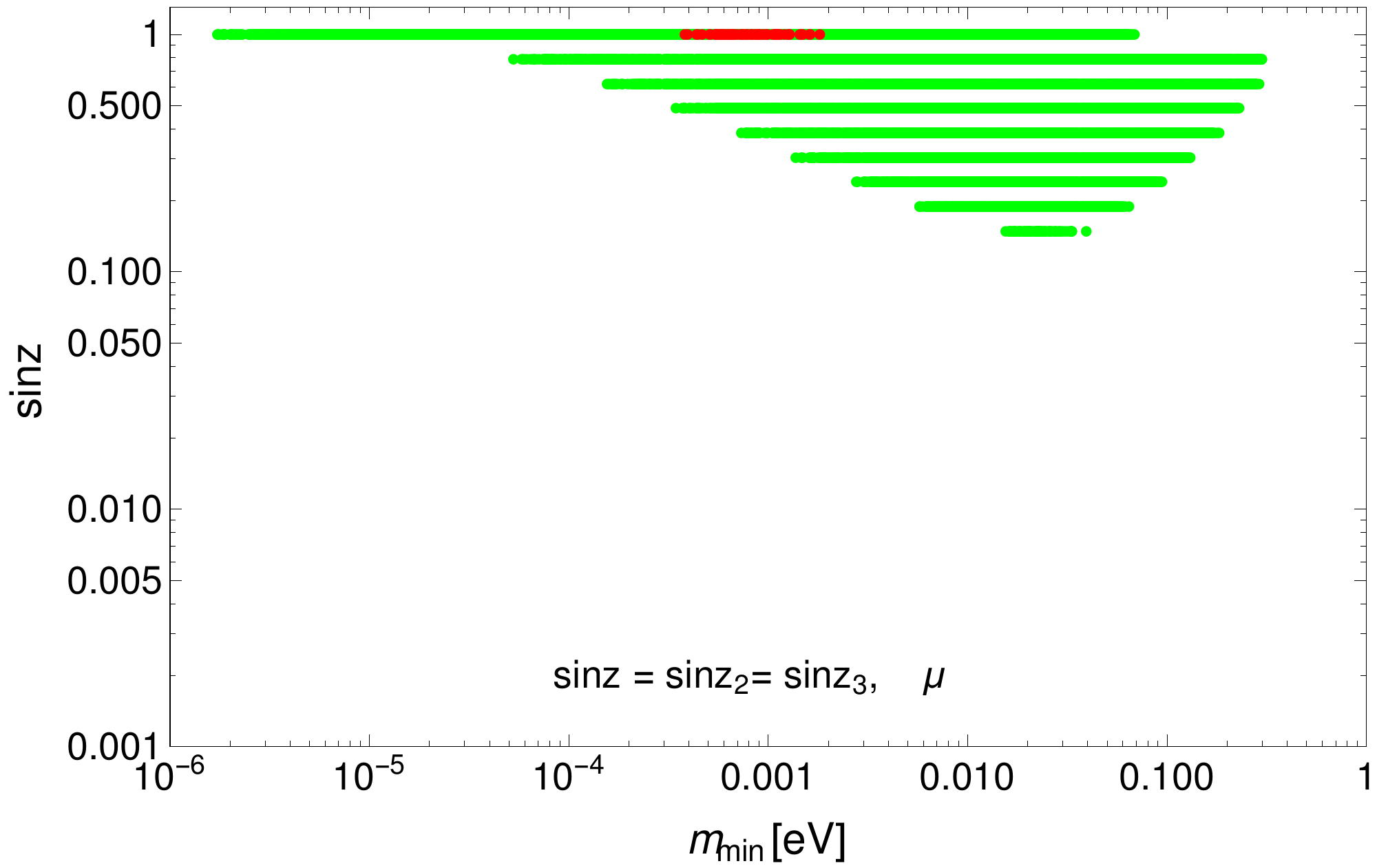}}}
    \qquad
{{\includegraphics[width=65mm]{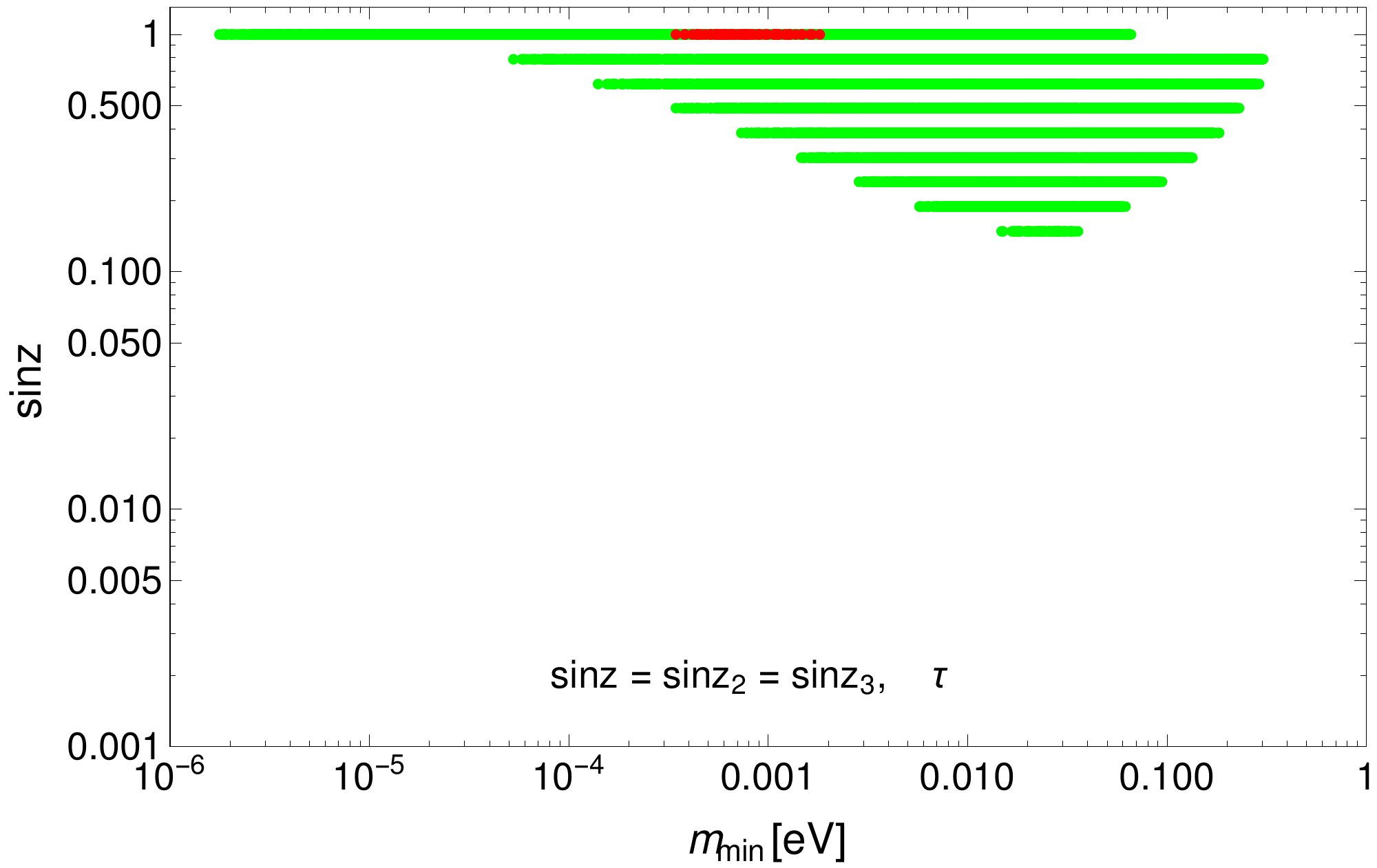}}}
\caption{\footnotesize{Allowed points in the plane of $\sin z$ versus $m_{\rm min}$ when $|m_{ee}|$ is maximised,
which give rise to $\epsilon_{1} > 10^{-8}$ $\epsilon^{l}, l=e ({\rm Fig}. a),\mu ({\rm Fig}. b),\tau ({\rm Fig}. c)$ . We assumed the
hierarchy to be IH and used two values $M_{1}$ $M_1=10^9$ GeV (Red) and $M_1=10^{10}$ GeV (Green) . We also set $\sin z_{2}$=$\sin z_{3}$=$\sin z$.} }
     \label{mmin_vs_resinz_IH_meemax}
\end{figure}

Note that the lower bound $M_{1}$ obtained here is two orders of magnitude larger than Davidson-Ibarra 
bound\,\cite{Davidson:2002qv}. We believe these occur due to the following two reasons: 
\begin{itemize}
\item The restricted choice of Majorana phases $\alpha_{1}$ and $\alpha_{2}$ obtained from the minimization of $|m_{ee}|$. 
\item The choice of $R$ is to be real.		
\end{itemize}
One way to lower this bound is to consider $R$ to be complex \cite{Davidson:2002qv}. However here we consider the alternative values of $\alpha_{1}$ and $\alpha_{2}$ by maximising the $|m_{ee}|$. These values of $\alpha_{1}$ and $\alpha_{2}$ are plotted in Fig.\,\ref{mmin_a1a2_max} as function of $m_{\rm min}$. The CP asymmetry parameter $\epsilon_{1}^{l}$ are computed using these values of $\alpha_1$ and $\alpha_2$ while all the other parameters are varied in the ranges mentioned previously. In Fig.\,\ref{mmin_vs_resinz_NH_meemax} and Fig.\,\ref{mmin_vs_resinz_IH_meemax} we plot the allowed values of $\sin z$ versus $m_{\rm min}$ which satisfy the constraint $|\epsilon_{1}^{l}| >10^{-8}$. Both for NH (Fig.\,\ref{mmin_vs_resinz_NH_meemax}) and IH (Fig.\,\ref{mmin_vs_resinz_IH_meemax}), we find the allowed values for $M_{1}=10^{9}$ GeV. This lower value is applicable only if $\sin z <0.25$ and $m_{\rm min}< 10^{-3}$ eV in the case of NH. On the other hand for IH, the lower value of $M_{1}=10^{9}$ GeV requires that $\sin z=1$ and $m_{\rm min}\simeq 10^{-3}$ eV, which is similar to the situation when $|m_{ee}|$ is minimized. 

\section{Flavoured CP-asymmetry with two right handed neutrinos}\label{CP-asy_2RHN}
It is possible to generate two independent mass square differences with only two non zero light neutrino masses, {\it i.e.,} we can set $m_{\rm min} \equiv 0$. This scenario can be achieved in the limit the heaviest right handed neutrino decouples. We implement this decoupling by setting the third row of the yukawa coupling matrix in Eq.\,\ref{casas} to be zero. Since one of the light neutrino masses is zero, one of the Majorana phases becomes unphysical and can be set equal to zero. Hence, the CP-asymmetries depend only on two phases $\delta$ and $\alpha_{1}$. We consider the cases of NH and IH seperately. 

\subsection{NH ($m_{1}=0$)}
In this case, the effective Majorana mass is
\begin{equation}
|m_{ee}|=m_{2}\cos^{2}\theta_{13} \sin^{2}\theta_{12} e^{-2 i \alpha_{1}}+m_{3}\sin^{2}\theta_{13} e^{2 i \delta}.
\label{mee_expr_a}
\end{equation}

As before, we need a cancellation between two terms to minimize $|m_{ee}|$, which leads to the condition $\alpha_{1} + \delta = (2n+1)\pi/2$. Because of the wide difference in the values of $\sin^{2}\theta_{12}$ and $\sin^{2}\theta_{13}$ the cancellation is never complete and the minimum of $|m_{ee}|$ in this case is of order $10^{-3}$ eV, as illustrated in Fig.\,\ref{mmin_vs_mee}.

The condition that the third row of the matrix $Y_{\nu}$ should consist of zeros leads to the following form of $R$,
	\begin{equation}
			R=
  			\begin{pmatrix}
    			0 & \cos z & \sin z \\
    			0 & -\sin z & \cos z \\
    			1 &  0  & 0
  			\end{pmatrix},
	\end{equation}
where we assume $z$ to be a real angle, as we did in three right handed neutrino case. The expression for the CP-asymmetry is
	\begin{eqnarray}
	\epsilon_{1}^{l}&=& -\frac{3 M_{1}}{16 \pi v^{2}}\,\, \frac{1}{m_{2}|R_{12}|^{2}+m_{3}|R_{13}|^{2}} \sqrt{m_{2}m_{3}}\left(m_{3}-m_{2}\right) \,[{\rm Im}(U_{l2}^{*}U_{l3})] \,R_{12}R_{13}.
	\label{cp_asy_2RHN_c}
	\end{eqnarray}	
	
\begin{figure} [h!]
	\centering
{{\includegraphics[width=70mm]{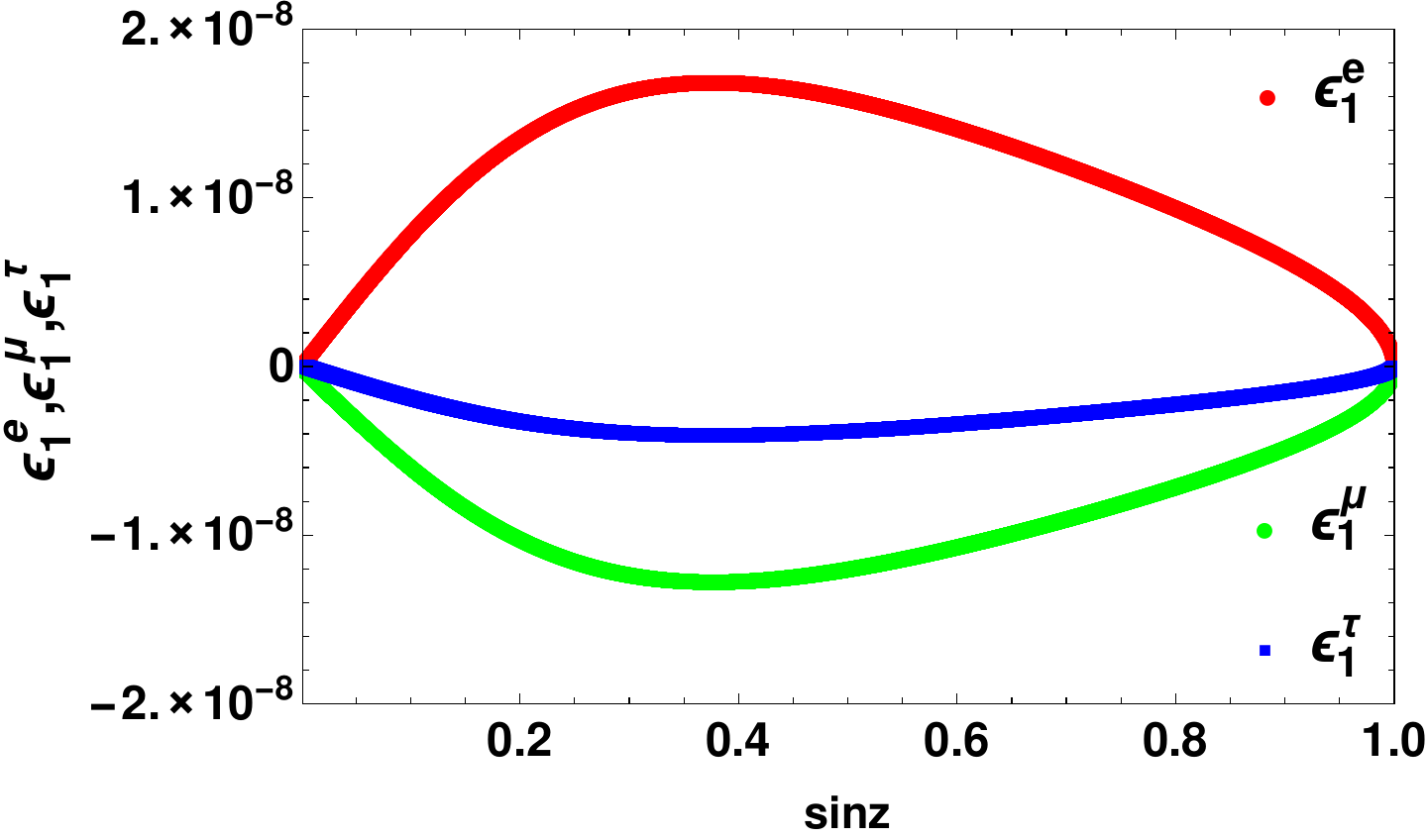}}}
 	     \qquad
{{\includegraphics[width=70mm]{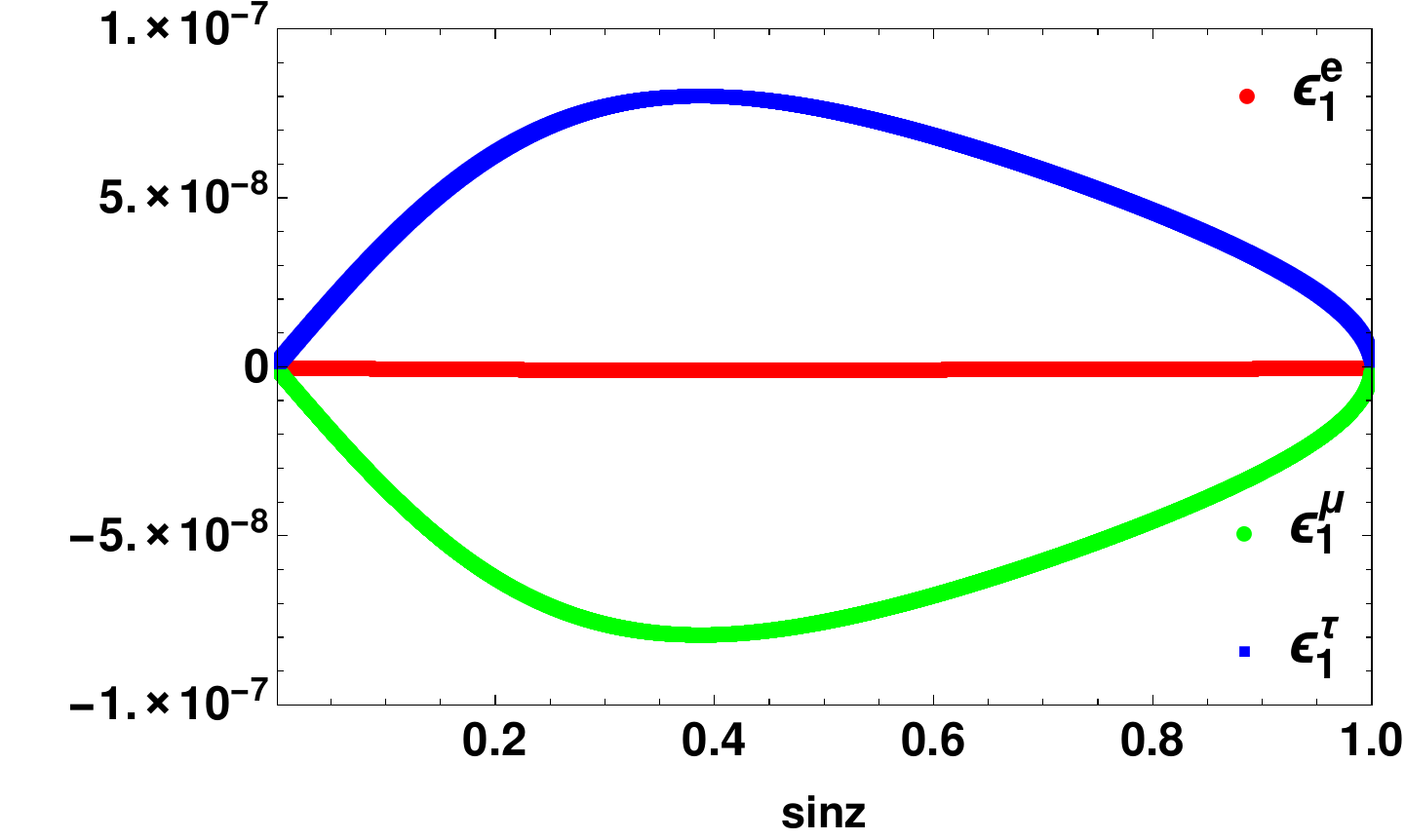}}}	
   \caption{\footnotesize{The CP-asymmetry corresponding to different flavours of leptons $\epsilon_{1}^{e}\, ,\,  \epsilon_{1}^{\mu}\,, \epsilon_{1}^{\tau}\,$. The inputs used are $M_{1}=10^{10}$ GeV,  $\delta = -\pi/2$ and the hierarchy is NH. (a) The value of the Majorana phase is $\alpha_{1}\simeq -\pi$, which minimizes $|m_{ee}|$. (b). The value of the Majorana phase is $\alpha_{1}\simeq -\pi/2$ (which maximizes $|m_ee|$.}}
	\label{Resinz_vs_eps_2RHN_min_max}
\end{figure}

In Fig.\,\ref{Resinz_vs_eps_2RHN_min_max} (left panel) we have plotted $\epsilon_{1}^{l}$ as a function of $\sin z$. We inputted $M_{1}=10^{10}$ GeV, $\delta=-\pi/2$ and $\alpha_{1} \simeq -\pi$, the Majorana phase obtained by minimizing $|m_{ee}|$. For this value of $M_{1}$, $|\epsilon_{1}^{l}| \sim 10^{-8}$ is possible. Lower values $M_{1}$ do not give rise to adequate leptogenesis. 
We also checked if smaller values of $M_1$ will be allowed if the Majorana phase is fixed by maximizing $|m_{ee}|$. In Fig.\,\ref{Resinz_vs_eps_2RHN_min_max} (right panel), we find that a value of $M_{1}=10^{10}$ GeV is required to obtain $\epsilon_{1}^{l} \simeq 10^{-8}$ for the values of the phases, $\delta=-\pi/2$ and $\alpha_{1} \simeq -\pi/2$. Since $\epsilon_{1}^{l} \propto M_{1}$ for $M_{1}<10^{10}$ GeV we cannot get $\epsilon_{1}^{l}>10^{-8}$.
	
\subsection{IH ($m_{3}=0$)}
In this case, the effective Majorana mass is
\begin{equation}
|m_{ee}|= m_{1} \cos^{2} \theta_{12} \cos^{2}\theta_{13}+ m_{2}\cos^{2}\theta_{13} \sin^{2}\theta_{12}\,\, e^{-2 i \alpha_{1}}.
\label{mee_expr_a}
\end{equation}

As before, we need a cancellation between two terms to minimize $|m_{ee}|$, which leads to the condition $\alpha_{1}  = (2n+1)\pi/2$. An exact cancellation is not possible because $m_{2}\gtrsim m_{1}$ and $\cos^{2}\theta_{12} \simeq 2 \sin^{2}\theta_{12}$. We are led to a lower limit on $|m_{ee}|$ of the order $10^{-2}$ eV, as illustrated in Fig.\,\ref{mmin_vs_mee}.

Once again we impose the condition that the third row of the matrix $Y_{\nu}$ should consist of zeros. This leads to the following form of $R$,
	\begin{equation}
			R=
  			\begin{pmatrix}
    			 \cos z & \sin z & 0 \\
    			-\sin z & \cos z & 0\\
    		       0  & 0 & 1
  			\end{pmatrix},
	\end{equation}
where we again assume $z$ to be a real angle. The expression for the CP-asymmetry is
	\begin{eqnarray}
	\epsilon_{1}^{l}&=& -\frac{3 M_{1}}{16 \pi v^{2}}\,\, \frac{1}{m_{1}|R_{11}|^{2}+m_{2}|R_{12}|^{2}} \sqrt{m_{1}m_{2}}\left(m_{2}-m_{1}\right) \,[{\rm Im}(U_{l1}^{*}U_{l2})] \,R_{11}R_{12}.
	\label{cp_asy_2RHN_c}
	\end{eqnarray}	
	
\begin{figure} [h!]
	\centering
{{\includegraphics[width=70mm]{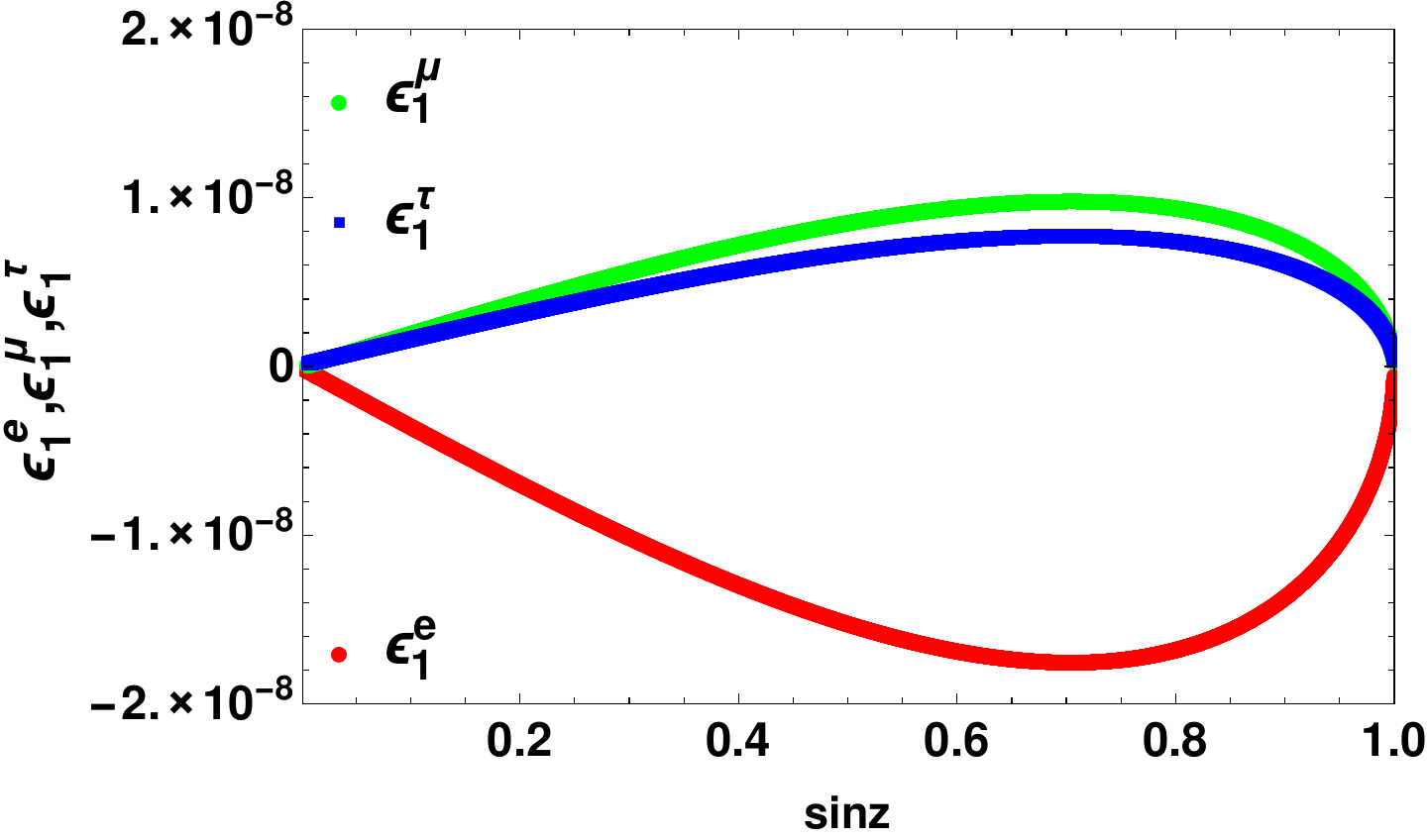}}}
 	     \qquad
{{\includegraphics[width=70mm]{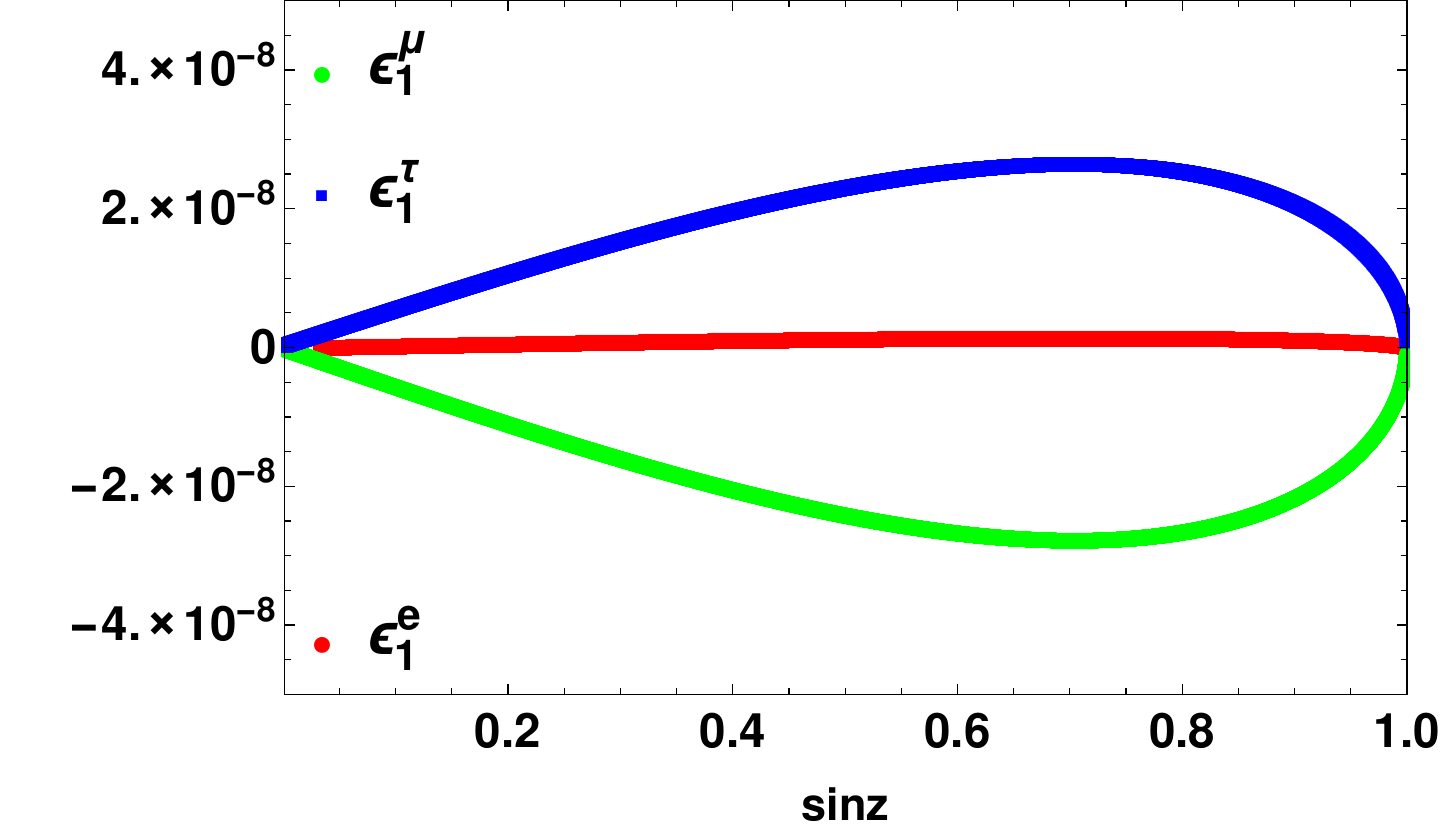}}}
\caption{\footnotesize{The CP-asymmetry corresponding to different flavours of leptons $\epsilon_{1}^{e}\, ,\,  \epsilon_{1}^{\mu}\,, \epsilon_{1}^{\tau}\,$. The inputs used are $\delta = -\pi/2$ and the hierarchy is IH. (a) The value of the Majorana phase is $\alpha_{1}\simeq -\pi$, which minimizes $|m_{ee}|$ and $M_1=10^{10}$ GeV. (b). The value of the Majorana phase is $\alpha_{1}\simeq -\pi/2$ (which maximizes $|m_ee|$ and $M_1=10^{12}$ GeV.}}
	\label{Resinz_vs_eps_2RHN_IH_min_max}
\end{figure}

As in the case of NH, we input $M_{1}=10^{10}$ GeV and $\delta=-\pi/2$. The value of $\alpha_{1}$ is taken to be $\simeq \pi/2$, which is the smallest value that minimizes $|m_{ee}|$. In Fig.\,\ref{Resinz_vs_eps_2RHN_IH_min_max}(left panel), we have plotted $\epsilon_{1}^{l}$ as a function of $\sin z$. Here again, a value of $M_{1}=10^{10}$ GeV is needed to obtain $|\epsilon_{1}^{l}| \sim 10^{-8}$. We also checked the parameter space for $\epsilon_{1}^{l} \geq 10^{-8}$ by maximizing $|m_{ee}|$. In Fig.\,\ref{Resinz_vs_eps_2RHN_IH_min_max}(right panel), we need $M_{1}=10^{12}$ GeV to obtain $\epsilon_{1}^{l} \simeq 10^{-8}$, for
$\delta=-\pi/2$ and $\alpha_{1} \simeq -\pi$. Further smaller values of $M_{1}$ cannot give rise to adequate leptogenesis.  Thus, in the case of $m_{\rm min}\rightarrow 0$, the lower limit on $M_{1}$, needed to generate adequate leptogenesis, rises to $10^{10}$ GeV.

\section{Conclusions}\label{conclusion}
In this work, we studied the correlation between the effective Majorana mass and the CP-asymmetry in case of flavour dependent leptogenesis. We minimize $|m_{ee}|$ to fix the Majorana phases of the PMNS matrix. We relate the neutrino yukawa matrix to light and heavy neutrino masses through Casas-Ibarra parameterization, which contains an unknown orthogonal matrix $R$. In general, this matrix can be complex and can give rise to additional CP violation. Here, we assumed this matrix to be real and parameterized in terms of a single angle $z$. We explored the possibility of obtaining adequate leptogenesis purely from the phases of PMNS matrix. Our analysis shows that a rather large right handed neutrino mass of $10^{10}$ GeV is needed to obtain such a result in the case of NH. For IH, $M_{1}=10^{9}$ GeV is possible provided $m_{3} \simeq 10^{-3}$ eV and $\sin z=1$. We relaxed the condition of minimization of $|m_{ee}|$ and fixed the values of Majorana phases by maximizing $|m_{ee}|$. With these values, adequate leptogenesis purely through the PMNS matrix is possible for $M_{1}=10^9$ GeV for both NH and IH. We also considered the case where the lightest neutrino mass is zero which corresponds to the case where one of the heavy right handed neutrino decouples. When the Majorana phases are fixed by the minimization of $|m_{ee}|$, the lower limit on $M_{1}$ is found to be $10^{10}\,(10^{11})$ GeV for NH (IH). If the Majorana phases are fixed by the maximization of $|m_{ee}|$, the lower limit on $M_{1}$ is found to be $10^{10}\,(10^{12})$ GeV for NH (IH). Thus our results show that the lower bound on $M_{1}$ in two right handed neutrino models is larger than the case of three right handed neutrinos. This could be due to lack of Majorana phases in case of former in comparison to the latter. 
 
\section{Appendix}
In this appendix, we discuss the form of the complex rotation matrix $R$ in the Casas-Ibarra parametrization, in the 
limit one of the light neutrino masses becomes zero and one of the heavy neutrino states decouples. The yukawa matrix
$(Y_\nu)_{il}$ in Eq.\,\ref{casas}, in general, is a $3 \times 3$ matrix. If the heavy neutrino state $M_3$ 
decouples, the elements of the $3$th row of this matrix vanish. From Eq.\,\ref{casas}, we obtain
\begin{equation}\label{casasappx}
(Y_{\nu})_{3l} = \frac{\sqrt{M_3}}{v} \left[ R_{31} \sqrt{m_1} (U^\dagger)_{1 l} +  R_{32} \sqrt{m_2} (U^\dagger)_{2 l} +
R_{33} \sqrt{m_3} (U^\dagger)_{3 l} \right].
\end{equation}
Suppose the light neutrino mass $m_1$ is set to zero, as we should for vanishing $m_{\rm min}$ in the case of NH. Then condition 
that the LHS of the above equation should vanish gives rise
to the constraints $R_{32} = 0 = R_{33}$. Orthogonality of $R$ implies that $R_{31} = 1$ and $R_{11} = 0 = R_{21}$. The four 
remaining elements of $R$, $R_{12}$, $R_{13}$, $R_{22}$ and $R_{23}$, form a $2 \times 2$ complex orthogonal matrix, defined
by one complex angle $z$. In the case of vanishing $m_{\rm min}$ for IH, we need to set $m_3=0$. It is easy to see from Eq.\,\ref{casasappx} that the decoupling of $M_3$ leads to the condition $R_{33} = 1$ which makes the third row and the third 
column of $R$ trivial. The upper $2 \times 2$ block of $R$ is a complex orthogonal matrix, once again parametrized by a single 
complex angle $z$.

The above argument can be extended to a general case. Suppose we want the heavy eigenstate with mass $M_i$ to 
decouple and we also want the light mass $m_j$ to be set to zero. The requirement that the $i$th row of $(Y_\nu)_{il}$
should vanish leads to the condition $R_{ij} = 1$ which means that the $i$th row and $j$th column of $R$ are trivial. 
The remaining four elements of $R$ then form a $2\times2$ complex orthogonal matrix parametrized by a single complex
angle $z$.

\end{document}